\documentclass[onecolumn,amsmath,amssymb,aps,showpacs,superscriptaddress,nofootinbib,noshowpacs]{revtex4}

\usepackage[dvips]{graphicx}
\usepackage[dvips]{color}
\usepackage{hyperref,breakurl,amsmath,amssymb,pst-node,eepic}
\usepackage{graphicx}
\usepackage{dcolumn}
\usepackage{bm}
\usepackage{hyperref}
\usepackage{float}

\begin{document}

\title{Excess reciprocity distorts reputation in online social networks}

\author{Giacomo Livan}
\author{Fabio Caccioli}
\author{Tomaso Aste}
\affiliation{Department of Computer Science, University College London, Gower Street, London WC1E 6EA, UK}
\affiliation{Systemic Risk Centre, London School of Economics and Political Sciences, Houghton Street, London WC2A 2AE, UK}

\date{\today}

\begin{abstract}
The peer-to-peer (P2P) economy relies on establishing trust in distributed networked systems, where the reliability of a user is assessed through digital peer-review processes that aggregate ratings into reputation scores. Here we present evidence of a network effect which biases digital reputation, revealing that P2P networks display exceedingly high levels of reciprocity. In fact, these are much higher than those compatible with a null assumption that preserves the empirically observed level of agreement between all pairs of nodes, and rather close to the highest levels structurally compatible with the networks' reputation landscape. This indicates that the crowdsourcing process underpinning digital reputation can be significantly distorted by the attempt of users to mutually boost reputation, or to retaliate, through the exchange of ratings. We uncover that the least active users are predominantly responsible for such reciprocity-induced bias, and that this fact can be exploited to obtain more reliable reputation estimates. Our findings are robust across different P2P platforms, including both cases where ratings are used to vote on the content produced by users and to vote on user profiles.
\end{abstract}

\maketitle

\section{Introduction}

The digital economy is increasingly self-organizing into a ``platform society'' \cite{vili,vandijck} where individuals exchange knowledge, goods, and resources on a peer-to-peer (P2P) basis. In recent years we have indeed witnessed how a number of well-established business-to-consumer sectors, such as the taxi and hotel industries \cite{proserpio}, have been disrupted by the emergence of the novel sharing economy P2P marketplaces.

P2P platforms rely on trust between their users. Trust is typically established by developing a reputation through digital peer-review mechanisms that allow users to rate their peers \cite{tadelis,vavilis}. Given the expected growth of the P2P paradigm, digital reputation will increasingly become central in our online lives, as it will determine access to substantial economic opportunities. Hence, it is crucial to ensure that digital peer-review systems produce reliable reputation scores.

Being decentralised, P2P systems are often thought to promote more economic freedom and democratisation. Yet, their current lack of regulation exposes them to a number of biases which can distort their functioning \cite{aral,watts,proserpio2,schweitzer}. Game theoretic considerations \cite{fehr,bolton,cimini}, and plenty of anecdotal evidence, suggest that users are often incentivised to reciprocate both positive and negative ratings.

In this paper  we show that P2P systems are indeed statistically characterized by excessive reciprocity, and that, on average, one reciprocated rating contributes to a user's reputation more than an unreciprocated one. The fact that reciprocity strongly affects reputation is rather relevant. Indeed, although it is true that the formation of social ties tends to be driven by homophily \cite{mcpherson}, which in turn might lead similar individuals to reciprocate more than they would do otherwise, it is also well documented that P2P platform users also exchange positive ratings to mutually boost reputation. The ``$5$ for $5$'' practice of Uber drivers and passengers, i.e. agreeing on exchanging $5$ star ratings at the beginning of a ride, is a common firsthand experience of such a practice \cite{dowd}, and similar phenomena have been reported in the interactions between eBay buyers and sellers \cite{resnick1,resnick2} and Airbnb hosts and guests \cite{fradkin}. Symmetrically, negative reciprocity due to mutually honest negative feedback needs to be distinguished from the retaliatory exchange of negative ratings. Therefore, our results show that a central issue in the design of P2P rating systems is that of discerning the information content of reciprocated ratings.

Reputation in online marketplaces is known to affect a buyer's willingness to pay \cite{resnick2}. In this respect, the exchange of ratings effectively introduces \emph{externalities} in P2P platforms \cite{nosko}, as they prevent users from making informed \emph{ex ante} decisions about their peers. Moreover, the anticipation of retaliatory behaviour can discourage users from providing negative feedback \cite{dellarocas}. In fact, it is well known that online ratings are often skewed towards positive values \cite{hu,proserpio2}. In summary, an excess of reciprocity may deteriorate the overall information content in P2P platforms and pose a threat to their fairness and transparency. 
We quantify the excess of reciprocity in P2P rating systems with respect to a range of null hypotheses, and to measure the impact it has in shaping online reputation. We do so by taking a deliberately reductionist approach. Namely, we investigate three case studies of P2P platforms with binary interactions, i.e. platforms whose users can either choose to endorse or reject their peers' activity. These environments represent a stylised template of the feedback systems underpinning P2P platforms, and allow for a parsimonious representation in terms of signed networks: A positive (negative) interaction between two platform users can be represented as a link carrying a positive (negative) sign between two nodes in a graph. Such systems have attracted considerable attention in the network literature \cite{leskovec1,leskovec2,bianconi}, as they offer a natural laboratory to test theories for systems with antagonistic interactions, such as social balance theory \cite{heider,altafini,thurner} and consensus formation \cite{altafini2}.

This network representation is particularly useful to quantify the statistical significance of reciprocity and its impact on user reputation. Indeed, the properties of multi-agent interacting systems can often be encoded into well defined {\em network motifs}. This, in turn, allows to quantify the statistical significance of empirically observed features by verifying whether certain motifs are still observed in null models in which the network topology is partially randomised. This approach has found successful applications in a great variety of fields, contributing to identifying relevant patterns, e.g., in the world trade web \cite{fagiolo} and in interbank credit networks \cite{squartini}.

We encode reciprocity as the fraction of mutual dyads in a network\cite{wasserman}, and compare the impact it has in shaping user reputation in the three networks we study as opposed to the one observed in two ensembles of null models. We build a first ensemble by partially randomizing the empirical networks with a link reshuffling procedure designed to preserve the reputation on each user at a predefined level of positive or negative reciprocity. We then proceed to investigate the role of user homophily, which, as mentioned above, may act as a natural source of reciprocity, especially in platforms where the nature of positive interactions is particularly friendly, e.g., where positive ratings are expressed by ``friending'' peers. This is particularly challenging with the data we employ, as they do not provide any categorisation of the nodes in terms of relevant features, which prevents any direct measurement of homophily. To this end, we devise a simple metric of preference similarity to quantify the propensity of pairs of nodes to agree in their endorsements or dislikes of other peers. We adopt such a metric as a proxy for homophily, and build a second null network ensemble through a link reshuffling procedure that preserves this quantity for each pair of nodes together with the reputation of each node. In conclusion, our work cannot explicitly distinguish between ``malicious'' reciprocity (as in the aforementioned $5$ for $5$ practice) and ``benign'' reciprocity, but provides statistically robust results by assessing the likelihood of generating empirically observed features of P2P platforms through partially randomised dynamics.

Within this framework, we quantify the role of rating reciprocity in distorting user reputations, and we eventually demonstrate that more reliable estimates of user reputation can be obtained by discounting reciprocated ratings from the least active users. As we will discuss in detail, the three case studies we analyse share a number of regularities despite the very different nature of the user interactions taking place in each of them.

\section{Results}

\subsection{Signed networks and reciprocity}
We analyse data from three P2P platforms: Slashdot, Epinions, and Wikipedia (see {\em Materials and Methods}). The data are freely available and can be downloaded from the Koblenz Network Collection repository \cite{koblenz}. We apply a filtering procedure to all three networks in order to discard the contributions from casual platform users and only retain the activity of actively engaged ones. In Table \ref{tab:network_stats} we provide details about the size and composition of such networks, and in Appendix \ref{robust} we provide detailed evidence that our results do not depend on the specificities of the filtering procedure.
\begin{table}[ht]
\centering
\begin{tabular}{|l|c|c|c|c|c|c|}
\hline
& $N_\mathrm{tot}$ & $N$ & $L^+$ & $L^-$ & $\xi^+$ & $\xi^-$ \\
\hline
Slashdot & $79,120$ & $4,611$ & $105,115$ & $29,190$ & $4.9 \times 10^{-3}$ & $1.4 \times 10^{-3}$ \\
\hline
Epinions & $131,828$ & $8,732$ & $425,377$ & $40,996$ & $5.6 \times 10^{-3}$ & $5.4 \times 10^{-4}$ \\
\hline
Wikipedia & $138,592$ & $5,538$ & $160,606$ & $29,327$ & $5.2 \times 10^{-3}$ & $9.6 \times 10^{-4}$ \\
\hline
\end{tabular}
\caption{\label{tab:network_stats} {\bf Network statistics.} The first column shows the total number of nodes $N_\mathrm{tot}$ in the original unfiltered networks. The other columns show the number of users $N$, the number of positive ($L^+$) and negative ($L^-$) ratings, and the sparsity $\xi^\pm$ (measured as the ratio between the number of existing links and the overall number of possible links, i.e. $\xi^\pm =  \sum_{i,j=1}^N \Theta (\pm A_{ij}) / (N(N-1))$) of the networks when restricted to a high participation core of actively engaged users with at least ten combined ratings (received and given).}
\end{table}

We represent a P2P platform of $N$ users exchanging positive and negative ratings as a signed network, i.e. a set of $N$ nodes described by a square $N \times N$ adjacency matrix $A$, whose entries are $A_{ij} = 1$ ($A_{ij} = -1$) if user $i$ has given a positive (negative) rating to user $j$, and $A_{ij} = 0$ if node $i$ has not rated node $j$. We say that a pair of positive (negative) links are reciprocated if $A_{ij} = A_{ji} = +1$ ($A_{ij} = A_{ji} = -1$). With this notation, one can introduce the number of unreciprocated positive and negative ratings received by a user $i$ (in the following $\Theta(x)$ denostes the step function such that $\Theta(x) = 1$ for $x> 0$ and $\Theta(x) = 0$ otherwise)
\begin{equation} \label{eq:link_types1}
\phi^+_i = \sum_{j=1}^N \Theta(A_{ji}) \left [ 1 - \Theta(A_{ij}) \right ] \ , \qquad
\phi^-_i = \sum_{j=1}^N \Theta(-A_{ji}) \left [ 1 - \Theta(-A_{ij}) \right ] \ ,
\end{equation} 
and the number of reciprocated positive and negative ratings received by a user $i$
\begin{equation} \label{eq:link_types2}
\gamma^+_i = \sum_{j=1}^N \Theta(A_{ji}) \Theta(A_{ij}) \ , \qquad 
\gamma^-_i = \sum_{j=1}^N \Theta(-A_{ji}) \Theta(-A_{ij}) \ .
\end{equation} 
In Appendix \ref{distribs} we provide details about the statistical properties of the above quantities. We then define $\Phi^+ = \sum_{i=1}^N \phi_i^+$ ($\Phi^- = \sum_{i=1}^N \phi_i^-$) and $\Gamma^+ = \sum_{i=1}^N \gamma_i^+$ ($\Gamma^- = \sum_{i=1}^N \gamma_i^-$) as the total number of positive (negative) ratings exchanged in the platform that belong to each category.

Within this context, we define positive (negative) reciprocity as the fraction of ratings $i \rightarrow j$ that have a matching rating $j \rightarrow i$ of the same sign, and we denote it as $\rho^+$ ($\rho^-$). With the above definitions we have 
\begin{equation} \label{eq:reciprocity_def}
\rho^+ = \frac{\Gamma^+}{L^+} \ , \qquad  
\rho^- = \frac{\Gamma^-}{L^-} \ ,
\end{equation}  
where $L^+ = \Phi^+ + \Gamma^+$ ($L^- = \Phi^- + \Gamma^-$) is the total number of positive (negative) links. From a network perspective, this definition is meaningful in the systems we work with as they are very sparse (see Table \ref{tab:network_stats}). In dense networks, where links are somewhat forced to reciprocate simply due to structural constraints, it could be replaced by the one introduced in \cite{garlaschelli}, which discounts density-related effects. This definition reads $\hat{\rho}^+ = (\rho^+ - \xi^+)/(1-\xi^+)$, where $\rho^+$ is as in \eqref{eq:reciprocity_def} and $\xi^+ = \sum_{i,j=1}^N \Theta(A_{ij})/(N(N-1))$ represents the density of the positive subnetwork. This definition straightforwardly generalizes to the negative subnetwork, and we report the corresponding values in Table \ref{tab:network_stats}. As it can be seen from the Table, all density values are of order $10^{-3}$ or $10^{-4}$, and therefore have a negligible impact on reciprocity.

\subsection{Reputation} 
Using the quantities introduced in Eqs. (\ref{eq:link_types1}) and (\ref{eq:link_types2}), we define the reputation of user $i$ as
\begin{equation} \label{eq:reputation}
R_i = \frac{L_i^+ - L_i^-}{L_i^+ + L_i^-} = \frac{\phi^+_i - \phi^-_i + \gamma^+_i - \gamma^-_i}{\phi^+_i + \phi^-_i + \gamma^+_i + \gamma^-_i} \ ,
\end{equation}
i.e. as the difference between the number of positive ($L_i^+ = \phi^+_i + \gamma^+_i$) and negative ($L_i^- = \phi^-_i + \gamma^-_i$) ratings received  normalised by the overall number of ratings received. The above definition of reputation is such that $-1 \leq R_i \leq 1$, where $R_i = 1$ ($R_i = -1$) for a user that has received positive (negative) ratings only.

We then define the average contributions to reputation associated with one unreciprocated or reciprocated positive (negative) rating, which we denote as $\lambda_\Phi^+$ and $\lambda_\Gamma^+$ ($\lambda_\Phi^-$ and $\lambda_\Gamma^-$), respectively. Recalling that $\Phi^\pm$ and $\Gamma^\pm$ indicate, respectively, the total numbers of unreciprocated and reciprocated positive/negative ratings in the networks, and introducing the total reputation in the network $R = \sum_{i=1}^N R_i$, we can write
\begin{equation*} \label{eq:rep}
R = \Phi^+ \lambda_\Phi^+ - \Phi^- \lambda_\Phi^- + \Gamma^+ \lambda_\Gamma^+ - \Gamma^- \lambda_\Gamma^- \ .
\end{equation*}

From the above equation one can obtain the following explicit expressions:
\begin{eqnarray} 
\label{eq:rep_production_rates}
&&\lambda_\Phi^+ = \frac{1}{\Phi^+} \sum_{i=1}^N \frac{\phi^+_i}{\phi^+_i + \phi^-_i + \gamma^+_i + \gamma^-_i} \qquad
\lambda_\Phi^- = \frac{1}{\Phi^-} \sum_{i=1}^N \frac{\phi^-_i}{\phi^+_i + \phi^-_i + \gamma^+_i + \gamma^-_i} \\ \nonumber
&&\lambda_\Gamma^+ = \frac{1}{\Gamma^+} \sum_{i=1}^N \frac{\gamma^+_i}{\phi^+_i + \phi^-_i + \gamma^+_i + \gamma^-_i} \qquad
\lambda_\Gamma^- = \frac{1}{\Gamma^-} \sum_{i=1}^N \frac{\gamma^-_i}{\phi^+_i + \phi^-_i + \gamma^+_i + \gamma^-_i} \ .
\end{eqnarray}

\subsection{Excess reciprocity}
We compare the reciprocity observed in the empirical networks with the one measured under two null assumptions designed to preserve the overall reputation landscape of the network: In both cases we reshuffle links in the network while preserving the numbers of positive/negative ratings received and given by each individual node (see {\em Materials and Methods}). We also introduce a positive/negative reciprocity target $\tau^\pm$ and we require the reshuffling algorithms to converge towards it by means of a cost function which depends on an ``intensity of choice'' parameter $\beta \geq 0$. When $\beta = 0$ the reshuffling procedure is fully random and not sensitive to the reciprocity target. On the other hand, for large values of $\beta$ the algorithms force the networks, compatibly with the aforementioned constraints, towards configurations that produce the desired level of reciprocity $\tau^\pm$ (see {\em Materials and Methods}).

We indicate the two null models we work with as NM1 and NM2, respectively, and we specify them as follows.
\begin{itemize}
\item NM1 produces randomised configurations of the empirical networks at a predefined positive (negative) reciprocity target $\tau^+$ ($\tau^-$) while preserving the reputation score $R_i$ (see Eq. (\ref{eq:reputation})) of each node $i$.
\item NM2 produces randomised configurations of the empirical networks at a predefined positive (negative) reciprocity target $\tau^+$ ($\tau^-$) while preserving both the reputation score of each node and the {\em preference similarity} of each pair of nodes.
\end{itemize}
We define the preference similarity of a pair of nodes $(i,j)$ as
\begin{equation} \label{eq:pref_sim}
S_{ij} = \sum_{\ell=1}^N A_{i\ell} A_{j\ell} \ .
\end{equation}
The above quantity measures the level of agreement between two nodes. Indeed, whenever nodes $i$ and $j$ both endorse or dislike a third peer $\ell$ (i.e. when $A_{i\ell} = A_{j\ell} = 1$ or $A_{i\ell} = A_{j\ell} = -1$) the above sum increases by one unity, whereas when the two nodes disagree (i.e. when $A_{i\ell} \neq 0$, $A_{j\ell} \neq 0$, and $A_{i\ell} = - A_{j\ell}$) it decreases by one unity. In this respect, we consider preference similarity as a reasonable proxy for the underlying level of homophily between pairs of nodes. Let us also remark that NM1 preserves preference similarity {\em globally}, i.e. it preserves the sum $S = \sum_{i <j} S_{ij}$. On the other hand, NM2 preserves preference similarity both locally and globally, and it also captures fairly well the empirical statistical properties of the preference similarity between pairs of nodes that share and do not share reciprocated relationships (see Appendix \ref{nm2}).

We first compare the empirically observed reciprocity levels with the ones measured in the above null models when carrying out the link reshuffling at $\beta = 0$, which produces maximally randomised counterparts of the reputation landscapes observed in the empirical networks, and therefore allows to measure a ``basal'' reciprocity $\rho_0^\pm$ that remains in the system due to its density and the constraints on reputation and preference similarity. Table \ref{tab:pos_reciprocity} shows that the positive reciprocity measured in the empirical networks is markedly over-expressed with respect to the positive basal reciprocity levels in both null models. Table \ref{tab:neg_reciprocity} reports the results we obtained for negative reciprocity. Slashdot, is the only network whose empirical negative reciprocity is over-expressed with respect to both null models. Indeed, while negative reciprocity in Epinions and Wikipedia is substantially over-expressed with respect to NM1, it is compatible with the one measured in NM2 in Epinions, and it is slightly under-expressed with respect to NM2 in Wikipedia.
\begin{table}[h!]
\centering
\begin{tabular}{|l|c|c|c|c|c|}
\hline
& & \multicolumn{2}{|c|}{Null model 1} & \multicolumn{2}{|c|}{Null model 2} \\
\hline
& $\rho^+$ & $\rho^+_0$ & $\rho^+_\mathrm{SAT}$ & $\rho^+_0$ & $\rho^+_\mathrm{SAT}$ \\
\hline
Slashdot & $41.3\%$ & $[2.28; 2.53]\%$ & $[46.6, 47.0]\%$ & $[4.81; 4.96]\%$ & $[42.4;42.6]\%$ \cr
\hline
Epinions & $42.4\%$ & $[2.33; 2.49]\%$ & $[48.2; 48.4]\%$ & $[4.58; 4.66]\%$ & $[43.1;43.2]\%$ \cr
\hline
Wikipedia & $17.6\%$ & $[3.01; 3.27]\%$ & $[36.8; 37.2]\%$ & $[3.49; 3.62]\%$ & $[25.4; 25.8]\%$ \cr
\hline
\end{tabular}
\caption{\label{tab:pos_reciprocity} {\bf Over-expression of positive reciprocity in P2P platforms.} Comparison between the positive reciprocity $\rho^+$ observed in the three networks we analyse and the $99\%$ confidence level intervals for the corresponding ``basal'' levels $\rho_0^+$ and saturation levels $\rho^+_\mathrm{SAT}$ obtained under a null hypothesis of random link rewiring constrained to preserve each user's reputation (null model 1), and a null hypothesis further constrained to also preserve the preference similarity of each pair of nodes (null model 2, see Eq. (\ref{eq:pref_sim})).}
\end{table}
\begin{table}[h!]
\centering
\begin{tabular}{|l|c|c|c|c|c|}
\hline
& & \multicolumn{2}{|c|}{Null model 1} & \multicolumn{2}{|c|}{Null model 2} \\
\hline
& $\rho^-$ & $\rho^-_0$ & $\rho^-_\mathrm{SAT}$ & $\rho^-_0$ & $\rho^-_\mathrm{SAT}$ \\
\hline
Slashdot & $15.9\%$ & $[0.35; 0.62]\%$ & $[25.6; 26.2]\%$ & $[9.05; 9.55]\%$ & $[20.2; 20.8]\%$ \cr
\hline
Epinions & $7.70\%$ & $[1.08; 1.42]\%$ & $[24.9; 25.3]\%$ & $[7.33; 7.90]\%$ & $[18.5;19.0]\%$ \cr
\hline
Wikipedia & $8.50\%$ & $[1.84; 2.45]\%$ & $[48.4; 49.1]\%$ & $[8.57; 9.18]\%$ & $[33.7; 34.2]\%$ \cr
\hline
\end{tabular}
\caption{\label{tab:neg_reciprocity} {\bf Over-expression of negative reciprocity in P2P platforms.} Comparison between the negative reciprocity $\rho^-$ observed in the three networks we analyse and the $99\%$ confidence level intervals for the corresponding ``basal'' levels $\rho_0^-$ and saturation levels $\rho^-_\mathrm{SAT}$ obtained under a null hypothesis of random link rewiring constrained to preserve each user's reputation (null model 1), and a null hypothesis further constrained to also preserve the preference similarity of each pair of nodes (null model 2, see Eq. (\ref{eq:pref_sim})).}
\end{table}

For large $\beta$, we can instead push the reshuffled networks towards targets higher than the reciprocity observed in the empirical networks. We find that all three platforms reach a saturation both in positive and negative reciprocity, i.e. the networks run out of links that can be used to reciprocate while still preserving each node's reputation and, in the case of NM2, local preference similarity. We report such values as $\rho^+_\mathrm{SAT}$ and $\rho^-_\mathrm{SAT}$ in Tables \ref{tab:pos_reciprocity} and \ref{tab:neg_reciprocity}, respectively. Both Slashdot and Epinions reach saturation shortly after the target $\tau^+$ exceeds the actual positive reciprocity $\rho^+$, i.e. the ratio $\rho^+_\mathrm{SAT} / \rho^+$ is only slightly larger than one, especially in NM2 where the local structure of preference similarity is kept intact. On the contrary, Wikipedia can sustain values of reciprocity much larger than $\rho^+$, i.e. the ratio $\rho^+_\mathrm{SAT} / \rho^+$ is larger than $2$ in NM1 and close to $1.5$ in NM1. We relate this to the different nature of the interactions. Indeed, interactions in Slashdot, where positive and negative links correspond to users tagging each other as ``friend'' or ``foe'', encourage backscratching and retaliatory behaviour, whereas a collaborative environment such as Wikipedia is subject to a different incentive structure. This picture is corroborated by the findings on negative reciprocity, where the ratio $\rho^-_\mathrm{SAT} / \rho^-$ increases substantially as progressing from Slashdot to Wikipedia. The general remark one can make from such results is that more polarised P2P environments are closer to their reciprocity saturation levels.

\subsection{Production of reputation through reciprocity}

We now ask whether reciprocity biases reputation in P2P systems, and, if so, to what extent. To this end, we have divided ratings into four categories: unreciprocated positive ratings, unreciprocated negative ratings, reciprocated positive ratings, and reciprocated negative ratings (see Eqs. (\ref{eq:link_types1}) and (\ref{eq:link_types2})). Unreciprocated ratings can be reasonably assumed to represent objective assessments, and their contribution to reputation can be thought of as a proxy of a user's ``true'' reputation. On the other hand, a fraction of the reciprocated ratings could be due to collusion and retaliation.

We compute the average contribution to reputation coming from ratings belonging to each of the four above categories. We do so by means of the quantities introduced in Eq. (\ref{eq:rep_production_rates}). Table IV reports the values of these quantities. In all cases we find that $\lambda_\Gamma^+ > \lambda_\Phi^+$, i.e. on average a reciprocated positive rating contributes more to total reputation than an unreciprocated one. We instead find mixed signatures in the case of negative ratings: only in Slashdot, where negative interactions are genuinely hostile (users labeling their peers as ``foes''), we observe $\lambda_\Gamma^- > \lambda_\Phi^-$, i.e. that reciprocated negative ratings play a larger role in damaging reputation than unreciprocated ones. We interpret this as a signature of retaliatory behaviour.
\begin{table*}[h!]
\centering
\begin{tabular}{|l|c|c|c|c|}
\hline
& $\lambda_\Phi^+$ & $\lambda_\Gamma^+$ & $\lambda_\Phi^-$ & $\lambda_\Gamma^-$  \\
\hline
Slashdot & $0.0323$ & $0.0355$ & $0.0340$ & $0.0522$ \\
\hline
Epinions & $0.0156$ & $0.0220$ & $0.0236$ & $0.0157$ \\
\hline
Wikipedia & $0.0270$ & $0.0325$ & $0.0362$ & $0.0321$ \\
\hline
\end{tabular}
\label{tab:rep_production}
\caption{{\bf Evidence that reciprocated ratings contribute more to reputation than unreciprocated ones.} $\lambda_\Phi^\pm$ denote the average contribution to reputation from a positive/negative unreciprocated rating, while $\lambda_\Gamma^\pm$ denote the average contribution from a positive/negative reciprocated rating.}
\end{table*}

\begin{figure}[h!]
\centering
\includegraphics[width=1.0\columnwidth]{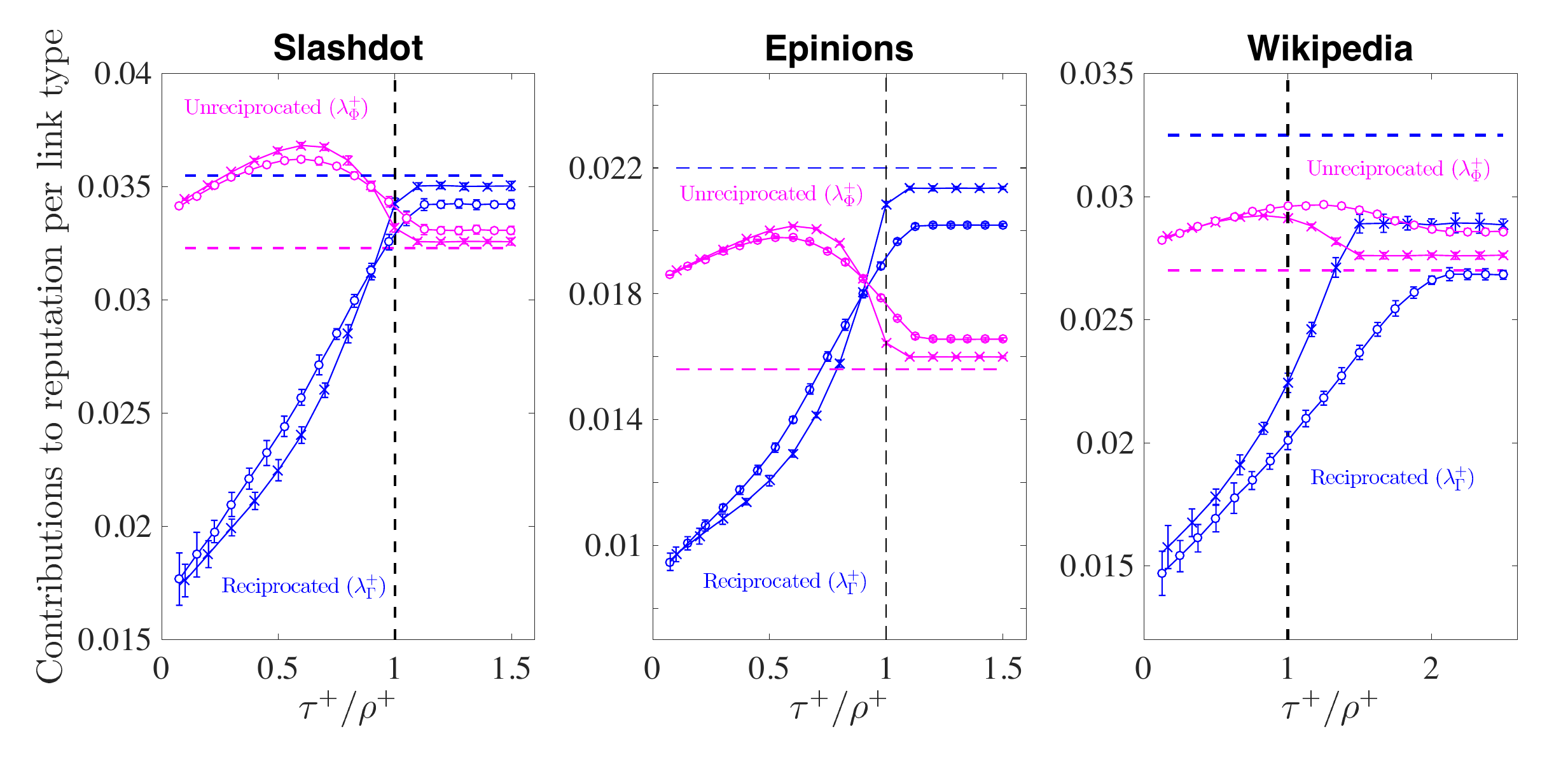}
\caption{\textbf{Demonstration that reputation is affected by the reciprocity bias.} Behaviour of the average contribution to reputation from unreciprocated positive ratings ($\lambda_\Phi^+$, pink) and reciprocated positive ratings ($\lambda_\Gamma^+$, blue) under two null hypotheses of random link rewiring designed to produce a predefined positive reciprocity target $\tau^+$. Circles refer to a null hypothesis constrained to preserve the reputation of each user (null model 1), while crosses refer to a null hypothesis further constrained to also preserve the preference similarity of each pair of nodes (null model 2, see also Eq. (\ref{eq:pref_sim})). The behaviour of $\lambda_\Phi^+$ and $\lambda_\Gamma^+$ is shown as a function of the ratio between the reciprocity target $\tau^+$ and the positive reciprocity $\rho^+$ measured in the actual platforms (column 1 of Table \ref{tab:pos_reciprocity}). Error bars correspond to $99\%$ confidence level intervals. Dashed lines correspond to the values of $\lambda_\Phi^+$ (pink) and $\lambda_\Gamma^+$ (blue) measured in the actual platforms (i.e. to the values reported in columns 1 and 2, respectively, of Table \ref{tab:rep_production}). The fact that the contribution from reciprocated (unreciprocated) activity in the actual platforms is systematically lower (higher) than under our null hypotheses highlights the existence of the reciprocity bias.}
\label{fig:lambda_high_beta}
\end{figure}

We test the statistical significance of the above findings by resorting again to our two null hypotheses based on constrained random link rewiring. Fig. \ref{fig:lambda_high_beta} shows the average contributions to reputation coming from positive reciprocated and unreciprocated ratings as functions of the ratio between the reciprocity target $\tau^+$ and the positive reciprocity $\rho^+$ of the empirical networks. A number of relevant results can be deduced from this Figure.

First, the behaviour of the two quantities as functions of $\tau^+$ is markedly different. The contribution from reciprocated positive ratings $\lambda_\Gamma^+$ is monotonically increasing, whereas the contribution from unreciprocated positive ratings $\lambda_\Phi^+$ attains a maximum in correspondence of a certain reciprocity target. In Slashdot and Epinions, such a value is around $50\%$ of the reciprocity observed in the empirical networks in both null models. Conversely, reciprocity in the Wikipedia network has to be increased with respect to its original level in order to reach the maximum contribution from unreciprocated activity. 

Second, throughout the whole range of reciprocity shown in Fig. \ref{fig:lambda_high_beta}, we find that the contribution to reputation from unreciprocated activity is under-expressed in the real systems. Symmetrically, the contribution from reciprocated ratings is systematically over-expressed. Both these conditions hold regardless of the specific null model (see Appendix \ref{link_signs} for a comparison with a different null model based on link and sign reshuffling), and hold for any value of the parameter $\beta$ (see Fig. \ref{fig:lambda_plus_app} in Appendix \ref{gen_nm1}). This highlights the existence of what we call \emph{reciprocity bias}: Reciprocated activity plays an exceedingly large role in shaping reputation at the aggregate level.

Third, in null models the relative importance between reciprocated and unreciprocated activity is reversed with respect to the one observed in the actual networks. As shown in Table IV, in the empirical networks one positive reciprocated rating always contributes more to reputation, on average, than an unreciprocated one (i.e. $\lambda_\Gamma^+ > \lambda_\Phi^+$). In contrast, under our null hypotheses the opposite holds over a wide range of the reciprocity target $\tau^+$. Namely, one has  $\lambda_\Gamma^+ < \lambda_\Phi^+$ almost up to the saturation threshold. Notably, both in Slashdot (for NM1) and Wikipedia one has $\lambda_\Gamma^+ < \lambda_\Phi^+$ even when the reciprocity target is kept equal to its empirical value, i.e. when $\tau^+ = \rho^+$ (dashed vertical lines in Fig. \ref{fig:lambda_high_beta}). This is a case where our null hypotheses entail the injection of a minimal amount of randomness into the system, as the only rewiring operations allowed are those that do not change reciprocity even at the local level. In the following we further demonstrate that P2P dynamics drive the networks towards very ``atypical'' states whose main features are not robust to small perturbations, and we will exploit this point to investigate possible prototypes of regulatory countermeasures to prevent users from building reputation through excessive reciprocity.

\subsection{Suppressing the reciprocity bias through random link elimination}

\begin{figure}[h!]
\centering
\includegraphics[width=1.0\columnwidth]{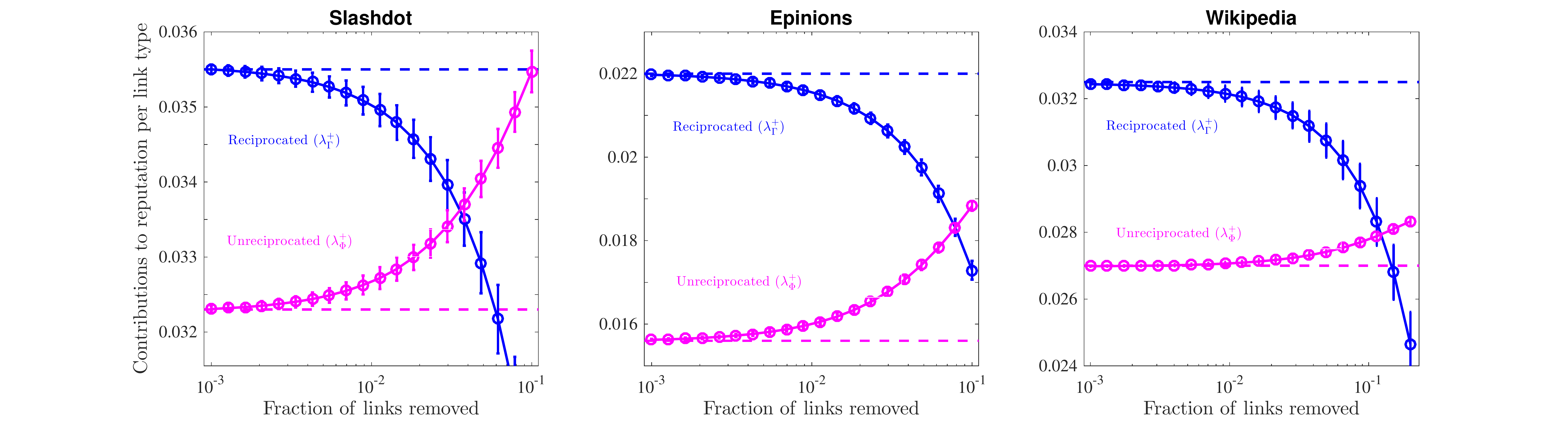}
\caption{\textbf{Demonstration that the elimination of a small fraction of ratings suppresses the reciprocity bias.} Solid lines show the average contribution to reputation from unreciprocated ($\lambda_\Phi^+$, pink) and reciprocated ($\lambda_\Gamma^+$, blue) positive links as a function of the fraction of reciprocated positive links removed from the network. Circles represent the behaviour of such quantities when a random node selection protocol is followed, i.e. nodes are chosen at random with uniform probability and reciprocated positive links between them, if any, are removed. Error bars represent $99\%$ confidence level intervals.}
\label{fig:link_removal}
\end{figure}

Fig. \ref{fig:link_removal} shows the behaviour of the average contribution to reputation from reciprocated ($\lambda_\Gamma^+$) and unreciprocated ($\lambda_\Phi^+$) positive ratings upon the removal of reciprocated ratings. Namely, we randomly select pairs of users and check whether a reciprocated rating between them exists. If so, we remove it. Notably, in Slashdot the deletion of $3\%$ of the reciprocated positive ratings (i.e. slightly more than $1.2\%$ of the overall positive ratings) is enough to make the contributions to reputation from reciprocated and unreciprocated ratings statistically compatible. The same result is achieved by removing roughly $8\%$ of the reciprocated positive ratings in Epinions, and roughly $11\%$ of the reciprocated positive ratings in Wikipedia (corresponding, respectively, to $3.3\%$ and $1.9\%$ of the overall positive ratings). Furthermore, one can also see from Fig. \ref{fig:link_removal} that statistical compatibility between $\lambda_\Gamma^+$ and $\lambda_\Phi^+$ as measured in the full networks and in the networks after the removal of a few ratings is lost extremely fast, i.e. by removing roughly $1\%$ of the reciprocated positive ratings.

Given the heavy tailed nature of the distributions of ratings given and received by each node (see Fig. \ref{fig:reputation_distr} in Appendix \ref{distribs}), a statistical feature common to many networked systems \cite{caldarelli}, the above protocol mostly targets and penalises users with a lower number of ratings, while leaving users with many reciprocated links relatively untouched. Although this might seem unfair at first, let us remark that the networks we analyse are high participation cores, where the contribution from casual platform users has already been filtered out. Moreover, such a protocol is meaningful from the viewpoint of user incentives: Indeed, a newcomer to a platform is more incentivised to reciprocate in order to boost her visibility in the network, whereas a high activity user with good reputation has little marginal gain from an additional rating. In this respect, removing ratings between low activity users is key to suppressing the reciprocity bias and improving the average quality of ratings (see Fig. \ref{fig:link_removal_app} in Appendix \ref{link_el}). 

The above exercise obviously neglects the complexities that translating such a procedure into an actual platform management policy would entail. Indeed, an explicit policy of random link removal of reciprocated links would discourage users from rating each other at all: no rating would be reciprocated (as any reciprocated pair would become liable to being removed), which in turn would discourage providing ratings in the first place. Yet, it is interesting to consider the above procedure as a thought experiment to measure the fragility of the reciprocity bias.

\section{Discussion}
The present paper provides a first systematic study of the network effects shaping digital reputation in P2P platforms. In this work we have tested the statistical significance of a number of empirical facts consistently observed in the three platforms we studied. We have done so by investigating a range of null models designed to preserve the individual reputation of each user and the preference similarities of pairs of users while probing different rating patterns that could have produced them. This effectively amounts to exploring ``alternate realities'' of P2P systems, while still keeping their heterogeneity fully intact at the level of individual users.

The overarching question we addressed in this framework is whether P2P platform users excessively engage in rating reciprocity in order to improve their reputation or to affect that of others. We do find that reciprocity, especially in the positive case, is substantially over-expressed with respect to null benchmarks. Moreover, we find that reciprocated ratings contribute more to reputation than unreciprocated ones. This is at odds with what we observe in the aforementioned null models, even when incorporating the tendency of users with similar like / dislike patterns to also like each other, where we always find that unreciprocated activity dominates the production of reputation. In other words, this shows that the same individual reputations are compatible with very different rating patterns between the users. In conclusion, the local structure of the networks is responsible for the distortions observed at the macroscopic level. 

The above point suggests that P2P systems exist in very peculiar states. Indeed, the contribution to reputation from reciprocated activity is systematically over-expressed with respect to all of the null hypotheses we consider, and a small random perturbation is enough to make unreciprocated activity the prevalent contribution. This point is evocative of other results concerning the beneficial effects of randomness in complex systems \cite{taleb,rapisarda1,rapisarda,mantegna,barabasi}, and suggests that an effective policy to prevent users from building reputation through excessive reciprocity in our simplified framework would be that of injecting a small amount of randomness into the system. We validated this hypothesis by carrying out a random link elimination procedure in the three networks we analysed, which shows that the removal of reciprocated links between users with a low number of ratings, hence highly incentivised to boost reputation, is most effective.

Our investigation highlights that interactions of a different nature (i.e., collaborative vs antagonistic) lead to different network signatures. Both in Slashdot and Epinions, suppressing reciprocity unquestionably makes unreciprocated ratings the prevalent contribution to reputation, up to a point (roughly corresponding to $50\%$ of the reciprocity observed in the actual networks, see Fig. \ref{fig:lambda_high_beta}) where the contribution from unreciprocated activity reaches a maximum. Conversely, and rather paradoxically, in the Wikipedia network reciprocity has to be increased with respect to its original level in order to reach the maximum contribution from unreciprocated activity. Indeed, Wikipedia reaches the highest average contribution to reputation from unreciprocated activity for reciprocity values higher than the one observed in the actual network. We speculate that this is due to the different outcomes that such networks aim to achieve. In essence, Wikipedia is a collaborative \emph{content-driven} environment whose users cooperate to the creation of a common good, i.e. knowledge. In contrast, Epinions and Slashdot have a more personal trait, as in both cases interactions are \emph{opinion-driven}: Users form relationships based on the endorsement or rejection of their peers' views. Our results suggest that an increase in reciprocity in a content-driven environment might lead to increased collaboration and, ultimately, to an improved quality of the ratings exchanged by the users. This aspect certainly deserves further attention through the analysis of other P2P networks.

Due to the lack of information about the users' identity and features, our paper does not investigate directly how homophily (i.e. the tendency to interact with similar individuals in social networks \cite{mcpherson,jackson}) might contribute to explain the exceedingly high levels of reciprocity and their impact on user reputation. Yet, the null network ensembles employed to carry out our statistical analyses are specifically designed to preserve the preference similarity structure of the networks, i.e. the tendency of certain pairs of users to endorse / reject the same content, which in itself represents a fairly close proxy for network homophily. In particular, we tested whether our conclusions are robust when compared to a null hypothesis that only preserves preference similarity at a global level as opposed to a more stringent null hypothesis designed to preserve preference similarity at the level of individual pairs of nodes. Interestingly, we found our results on positive excess reciprocity and its impact on reputation to be largely independent of the particular null assumption, i.e. the basal positive reciprocity levels and the contributions to reputation (see Fig. \ref{fig:lambda_high_beta}) observed when keeping the local preference similarity structure intact are not significantly different from those observed when this is not preserved. This suggests that homophily alone would not be able to justify the high levels of positive reciprocity we empirically observe and their impact on user reputation. On the other hand, preserving the local similarity structure essentially captures most of the empirical positive reciprocity. Indeed, although not statistically compatible, the saturation reciprocity levels measured when accounting for preference similarity are remarkably close to the empirically observed ones in Slashdot and Epinions. This suggests that homophily drives the growth of opinion-driven platforms in such a way as to maximize positive reciprocity. Symmetrically, the basal levels of negative reciprocity measured when accounting for homophily essentially capture the empirical levels in Epinions and Wikipedia, i.e. the exchange of negative ratings cannot be distinguished from a partially randomized dynamics when negative interactions lack an explicitly hostile trait.

The systems we study in this paper are simpler than the most popular P2P platforms where users build a peer-review based reputation, such as Uber and Airbnb. Yet, they retain the full complexity of those richer environments, both in terms of interaction patterns and user heterogeneity. It is precisely because of such a ``stylised-yet-complex'' nature that we chose signed networks as templates for P2P systems. In this respect, our work advances the existing literature on reciprocity and reputation in online environments, where most empirical results are tied to the specificities of a particular platform or rating system, and therefore lack generality. Our work provides a ``one-fits-all'' network methodology that can be used as soon as user interactions in a platform can be classified as either positive or negative, which can be achieved quite easily for the most common types of online feedback (e.g., by thresholding in the case of graded scales and via sentiment analysis in the case of textual ratings). The potential of such a reductionist approach is highlighted by the consistency of the regularities that we detect in the three platforms we analyse, despite the vastly different natures of the user interactions that characterise them. For instance, we observe qualitatively similar patterns of reputation formation in Slashdot, where votes are expressed directly on a whole user profile, and in Wikipedia, where edits express indirect votes on the quality of the content produced by others.

Our analyses show that P2P rating systems are plagued by biases. We have shown that the most widely adopted reputation metrics, i.e. those based on naive rating aggregation, are particularly vulnerable to distortions, which we have related with the presence of non-trivial network effects and motifs. In this respect, our work highlights the yet largely untapped potential that network science applications have in the digital economy domain, and suggests that novel, network-based, notions of reputation could be the way to ensure the fairness that P2P systems promise to deliver.

\section{Materials and methods}
\subsection{Data}
The data we analyze belong the the following platforms:
\begin{itemize}
\item {\bf Slashdot}, a website whose users post, read and comment news about science and technology. The data we analyse pertain to the Slashdot Zoo, i.e. the set of friendship/enmity relationships that Slashdot users form. Namely, users can label each other as ``friend'' (``foe'') in order to endorse (oppose) opinions and activity on the platform.
\item {\bf Epinions}, a platform for crowdsourced consumer reviews whose users exchange insight about a variety of products. Based on their review history, users can express trust/distrust relationship to each other.
\item A collection of actions from $563$ {\bf Wikipedia} articles about politics, where each interaction between users (such as co-edits, antagonistic edits, reverts, restores, etc.) was interpreted as positive or negative value depending on its nature.
\end{itemize}

A substantial portion of the users in the above networks are casual users who do not interact frequently with the platform. In fact, $32\%$ of Slashdot users, $47\%$ of Epinions users, and $49\%$ of Wikipedia users have either given or received just one rating. We therefore proceed to filter the noisy contribution from casual platform users, in order to ensure that reputation scores are computed from the ratings of actively committed users. The above network datasets do not provide information about possible repeated interactions between the users, which prevents from applying network statistical validation techniques that explictly account for the heterogeneity in user activity (see, e.g., \cite{mantegna}). Hence, we choose to restrict the networks to a high participation core of actively engaged users with a number of total ratings (received and given) equal or higher than a threshold $t$. Clearly, after the deletion of these nodes, some other nodes might result to be disconnected in the restricted network, as there is no guarantee that the $t$ (or more) received / given ratings are exchanged with other nodes within the core. We therefore carry out a second filtering in order to remove disconnected nodes. This operation leaves us with the networks whose statistical properties are described in Table \ref{tab:network_stats}. In Appendix \ref{robust} we provide evidence that our main results are consistent across different thresholds.

\subsection{Null models} 
\label{sec:null_mods}
In order to assess the statistical significance of the features observed in the empirical networks, we define two ensembles of null network models (labeled as NM1 and NM2 in the following) that depend on two parameters. Namely, we define a positive reciprocity target $\tau^+$, and we introduce a cost function $H(\tau^+) =  \left [ L^+ ( \rho^+ - \tau^+ \right ) ]^2$ to measure the distance between the current positive reciprocity in the network and the target (the following can be straightforwardly generalized to the case of negative reciprocity). Starting from the empirical networks, we perform rewiring operations in order to decrease the cost function's value, i.e. to make the networks' reciprocity converge to the predefined target. We do so in a probabilistic manner: we iteratively propose random rewiring operations and we accept them with probability
\begin{equation} \label{eq:rewiring_prob}
p(\beta,\tau^+) = \frac{e^{- \beta \Delta H(\tau^+)}}{1 + e^{- \beta \Delta H(\tau^+)}} \ ,
\end{equation}
where $\Delta H(\tau^+)$ measures the change in cost that would be achieved upon accepting the rewiring move (i.e. the difference between the cost function after and before the rewiring), and $\beta \geq 0$ is an ``intensity of choice'' parameter that determines the rewiring procedure's responsiveness to changes in cost: When $\beta = 0$ the above probability is equal to $1/2$, which amounts to a fully random rewiring (independently of the target $\tau^+$), whereas when $\beta$ is large rewiring moves that cause an increase (decrease) in cost get rejected (accepted) with probability close to one.

The link rewiring works as follows. In NM1:
\begin{itemize}
\item Two pairs of distinct nodes $(i,k)$ and $(j,\ell)$ connected by two positive links (i.e. $A_{ik} = A_{j\ell} = 1$), and such that $A_{i\ell} = A_{jk} = 0$, are chosen at random.
\item The change $\Delta H (\tau^+)$ in cost function that would be attained by disconnecting the existing links $i \rightarrow k$ and $j \rightarrow \ell$ and replacing them with links $i \rightarrow \ell$ and $j \rightarrow k$ (i.e. setting $A_{ik} = A_{j\ell} = 0$ and $A_{i\ell} = A_{jk} = 1$) is computed.
\end{itemize}
In NM2:
\begin{itemize}
\item Two triplet of nodes $(i,j,k)$ and $(h,g,\ell)$ connected by at least two pairs of positive links (i.e. such that $A_{ik} = A_{jk} = 1$ and $A_{h\ell} = A_{g\ell} = 1$), and such that $A_{i\ell} = A_{j\ell} = A_{hk} = A_{gk} = 0$, are chosen at random.
\item The change $\Delta H (\tau^+)$ in cost function that would be attained by disconnecting the existing links $i \rightarrow k$, $j \rightarrow k$ and replacing them with links $i \rightarrow \ell$ and $j \rightarrow \ell$, and by disconnecting the existing links $h \rightarrow \ell$, $g \rightarrow \ell$ and replacing them with links $h \rightarrow k$ and $g \rightarrow k$ (i.e. setting $A_{ik} = A_{jk} = A_{h\ell} = A_{g\ell} = 0$ and $A_{i\ell} = A_{j\ell} = A_{hk} = A_{gk} = 1$) is computed.
\end{itemize}
Then, for both null models:
\begin{itemize}
\item With the probability $p(\beta,\tau^+)$ in Eq. (\ref{eq:rewiring_prob}) the above rewiring moves are accepted and carried out. With probability $1-p(\beta,\tau^+)$ the rewiring moves are rejected and all links are kept as they are.
\item The above operations are repeated until a steady state is reached, which in the long run is ensured by the probabilistic rule in Eq. (\ref{eq:rewiring_prob}), where $\beta$ plays the role of an inverse temperature in a physical system (see Appendix \ref{convergence}).
\end{itemize}

In Fig. \ref{fig:rewiring_moves} we sketch the two fundamental rewiring moves of the above null models, and we report two examples in which they cause the overall positive/negative reciprocity of the network to decrease ($\Delta \rho^\pm < 0$) and to increase ($\Delta \rho^\pm > 0$).
\begin{figure}[ht]
\centering
\includegraphics[width=0.8\columnwidth]{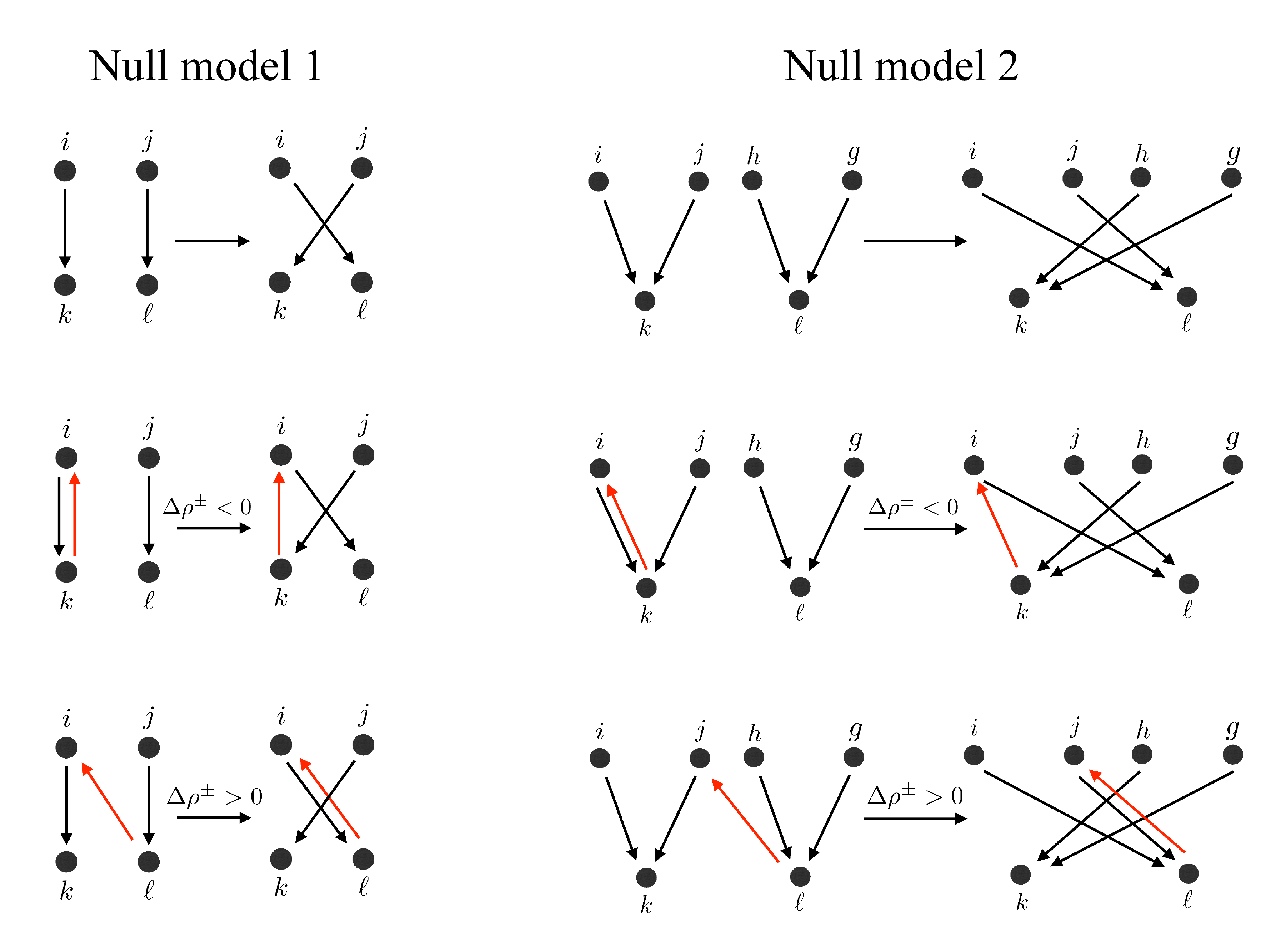}
\caption{\textbf{Rewiring moves of the null models.} On the top of the Figure we show the fundamental rewiring moves of the two null models we consider. In the central and bottom part of the Figure we also show examples where the rewiring moves contribute to decrease and increase the overall reciprocity, respectively. We highlight the links responsible for such increase / decrease in red.}
\label{fig:rewiring_moves}
\end{figure}

The above procedures are reminiscent of the directed configuration model from the literature on complex networks \cite{newman,dorogovtsev}, and consist in randomly redirecting ratings, hence destroying correlations between raters and ratees, while preserving both the reputation of each node and the system's heterogeneity at a fixed level of reciprocity. In fact, the above rewiring procedures preserve the number of positive/negative ratings received and given by each node $i$, i.e. they preserve the sums $\phi_i^+ + \gamma_i^+$ and $\phi_i^- + \gamma_i^-$, although, crucially, the individual values of $\phi_i^\pm$ and $\gamma_i^\pm$ are in general changed. Thus, according to Eq. (\ref{eq:reputation}) our rewiring procedures keep the reputation $R_i$ of each node intact.

\appendix

\section{Distributions of ratings and reputations}
\label{distribs}

\begin{figure*}[h!]
\centering
\includegraphics[width=\textwidth]{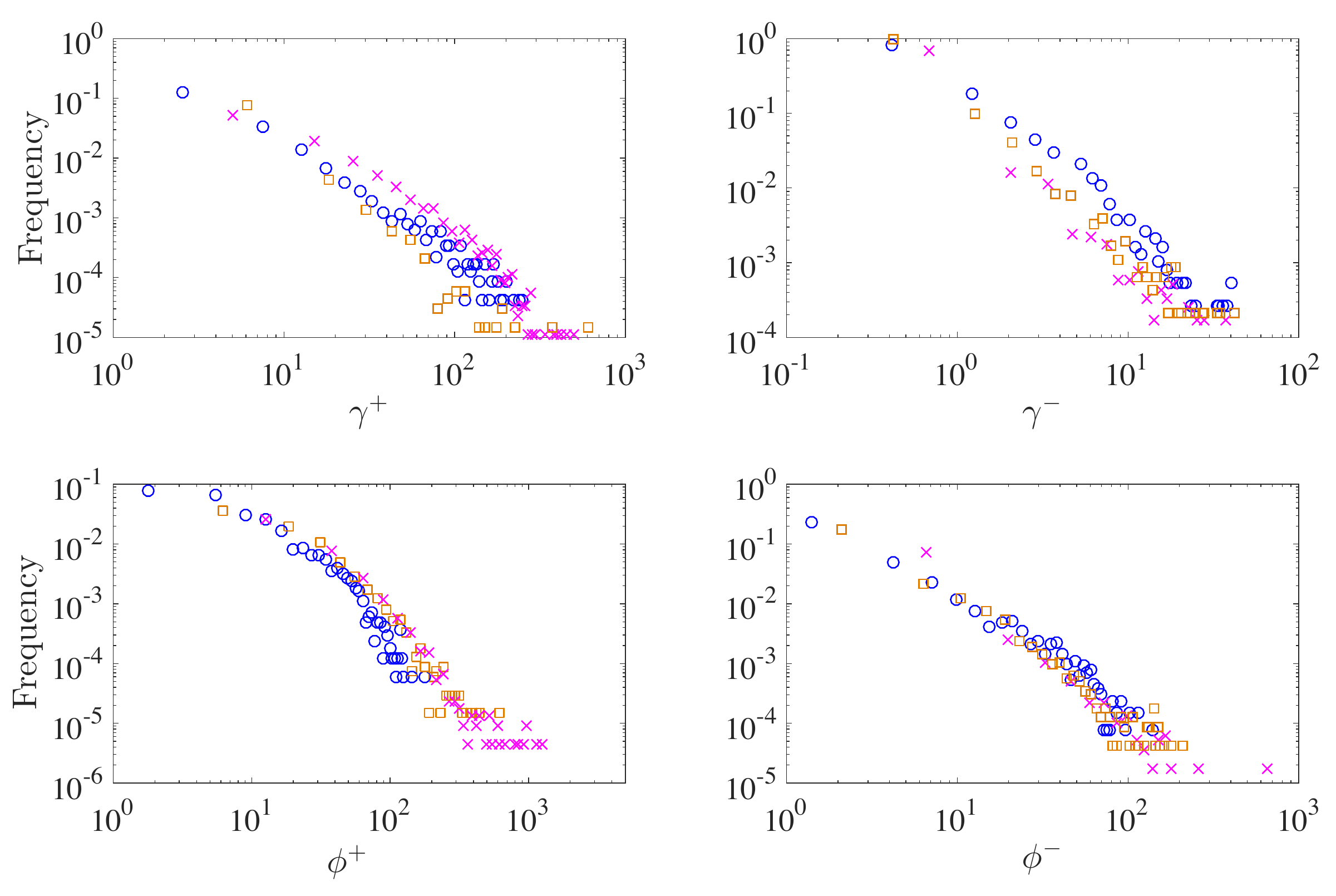}
\caption{{\bf Distributions of the number of positive/negative reciprocated and unreciprocated ratings received by each user}. Empirical distributions of the reciprocated positive ratings $\gamma^+$ (top-left panel), reciprocated negative ratings $\gamma^-$ (top-right panel), unreciprocated positive ratings $\phi^+$ (bottom-left panel), unreciprocated negative ratings $\phi^-$ (bottom-right panel) received by each user. Such quantities correspond to the ones introduced in Eqs. (1) and (2). In all plots blue circles represent Slashdot data, pink crosses represent Epinions data, and orange squares represent Wikipedia data.}
\label{fig:phi_gamma}
\end{figure*}

\begin{figure*}[h!]
\centering
\includegraphics[width=\textwidth]{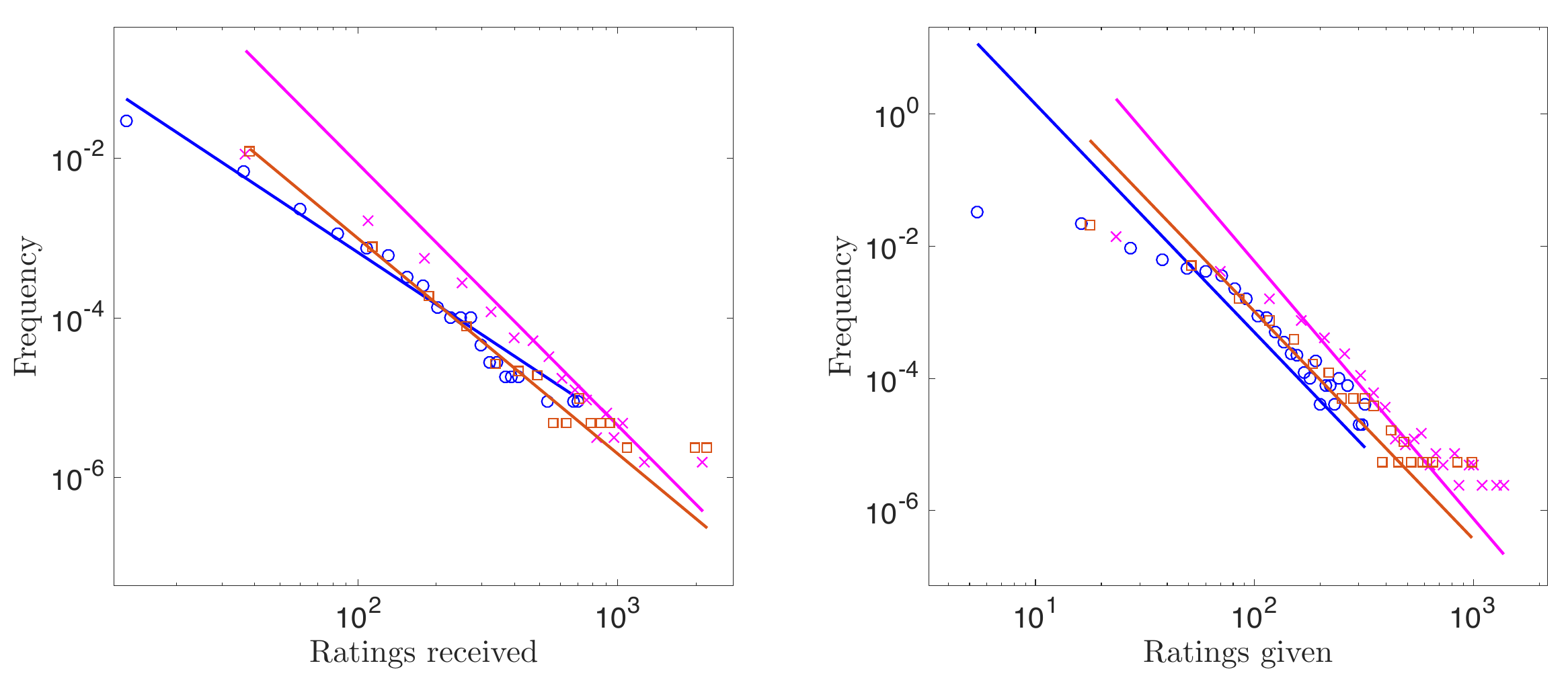}
\caption{{\bf Distribution of the ratings received and given by each user}. Empirical distribution of the ratings received (left panel) and given (right panel) by each node, plotted in log-log scale. Blue circles represent Slashdot data, pink crosses represent Epinions data, and orange squares represent Wikipedia data. Solid lines (of the same colours) show power law of the form $p(x) \sim x^{-\alpha}$ calibrated on the data with the method introduced in \cite{clauset}. For the distribution of received ratings we find $\alpha = 2.16$ (Slashdot), $\alpha = 3.28$ (Epinions), and $\alpha = 2.70$ (Wikipedia). In the case of given ratings we find $\alpha = 3.45$ (Slashdot), $\alpha = 3.89$ (Epinions), and $\alpha = 3.45$ (Wikipedia).}
\label{fig:ratings_distr}
\end{figure*}

As it is often the case in networked systems (see, e.g., \cite{caldarelli}), links are not distributed homogeneously between the nodes of the three platforms we analyse. Fig. \ref{fig:phi_gamma} shows the distributions of the four quantities we introduce in Eqs. (1) and (2), i.e. the number of reciprocated positive ratings ($\gamma^+$), reciprocated negative ratings ($\gamma^-$), unreciprocated positive ratings ($\phi^+$), and unreciprocated negative ratings ($\phi^-$) received by each user in the platform. Upon aggregation, these four quantities amount to the number of ratings received by each user (i.e. $\gamma^+ + \gamma^- + \phi^+ + \phi^-$), whose empirical distribution is shown in the left panel of Fig. \ref{fig:ratings_distr} together with the empirical distribution of the number of ratings given by each user to other peers in the platform. As it can be seen, all of these distributions are heavy tailed, and in Fig. \ref{fig:ratings_distr}
 we also report a calibrated probability distribution function of the form $p(x) \sim x^{-\alpha}$. In all cases we obtain values such that $2 < \alpha < 4$ (see the caption in Fig. \ref{fig:ratings_distr} for further details).


\begin{figure*}[h!] 
\centering
\includegraphics[scale=0.65]{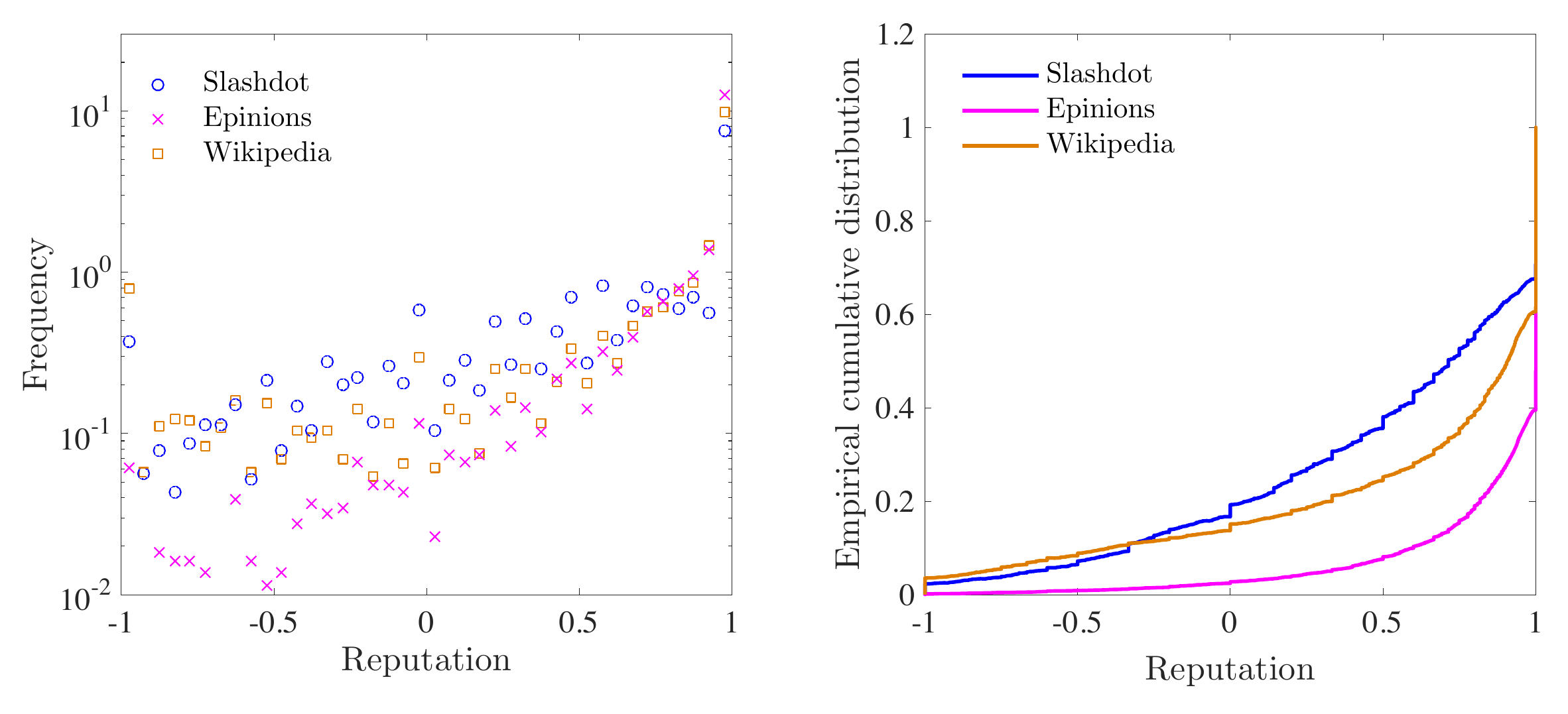}
\caption{{\bf Distributions of reputations.} The left panel shows the empirical probability density functions of the reputations in the three platforms we analyze, while the right panel shows the corresponding cumulative distributions. Reputations are computed as per Eq. (3).}
\label{fig:reputation_distr}
\end{figure*}

In Fig. \ref{fig:reputation_distr} we report the empirical probability densities and the cumulative distributions of the reputations, computed as per Eq. (3), of all users in the platforms we analyse. As a straightforward consequence of the marked prevalence of positive ratings (see Table I), a substantial fraction of users have reputations in the upper end of the spectrum. In particular, $35.9\%$ of Slashdot users, $58.1\%$ of Epinions users, and $45.1\%$ of Wikipedia have an ``immaculate'' reputation $R = 1$. This is clearly visible from the jump at $R = 1$ in the right panel in Fig. \ref{fig:reputation_distr}.

\section{Convergence of rewiring procedure}
\label{convergence}

\begin{figure*}[h!]
\centering
\includegraphics[scale=0.5]{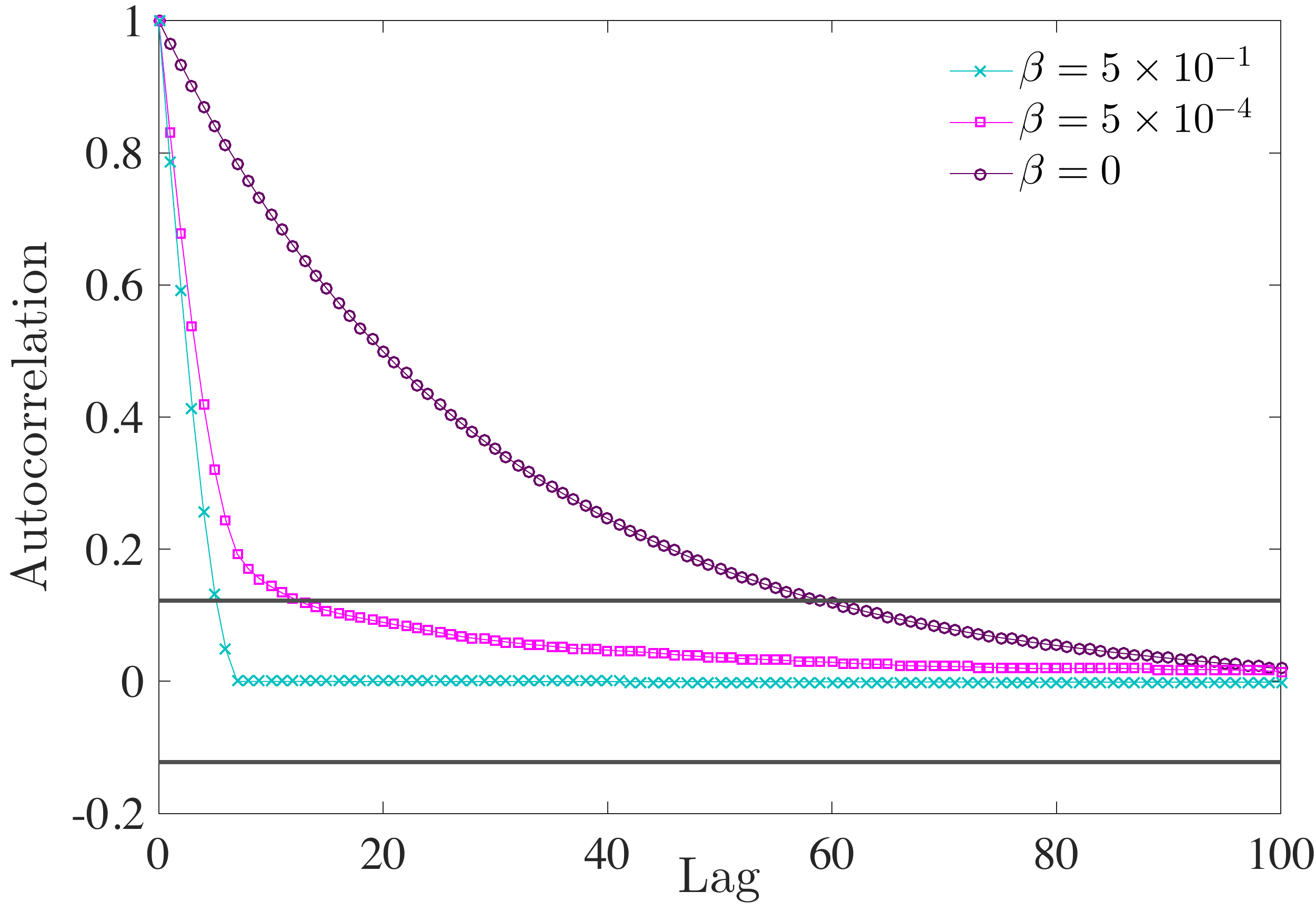}
\caption{{\bf Convergence of the rewiring procedure to a stationary state.} Circles, squares and crosses denote the autocorrelation function of the positive reciprocity in Slashdot as measured during the rewiring procedure described above. Different curves refer to different values of $\beta$ (shown in the legend), and in all cases the positive reciprocity target $\tau^+$ is set to $80\%$ of the positive reciprocity $\rho^+$ measured in the empirical network. Each time lag represents $10^3$ attempted rewiring moves. The solid horizontal lines denote the $95\%$ confidence level interval obtained under a null hypothesis of no autocorrelation. As it can be seen, in each case the rewiring procedure reaches a stationary state, albeit with different speeds. Indeed, convergence in the completely randomized case ($\beta = 0$) is much slower than in more ``selective'' cases characterized by higher values of $\beta$.}
\label{fig:autocorr}
\end{figure*}

As discussed in the main paper, the rewiring operations we carry out in order to sample configurations from our null model ensembles are repeated until a steady state is reached, which in the long run is ensured by the probabilistic rule in Eq. (5), where $\beta$ plays the role of an inverse temperature in a physical system. In Fig. \ref{fig:autocorr} we provide evidence that this is indeed the case by showing the autocorrelation function of the positive reciprocity measured in Slashdot as the rewiring procedure of null model 1 is carried out.

\section{General results for reputation in null model 1}
\label{gen_nm1}
\begin{figure*}[h!]
\centering
\includegraphics[width=\textwidth]{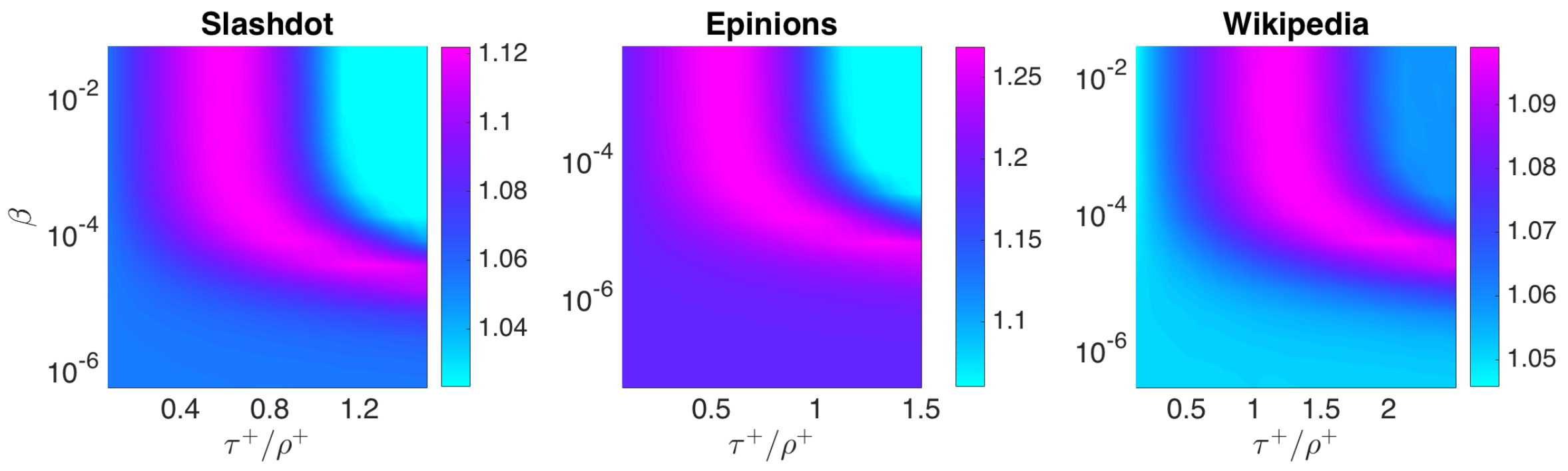}
\includegraphics[width=\textwidth]{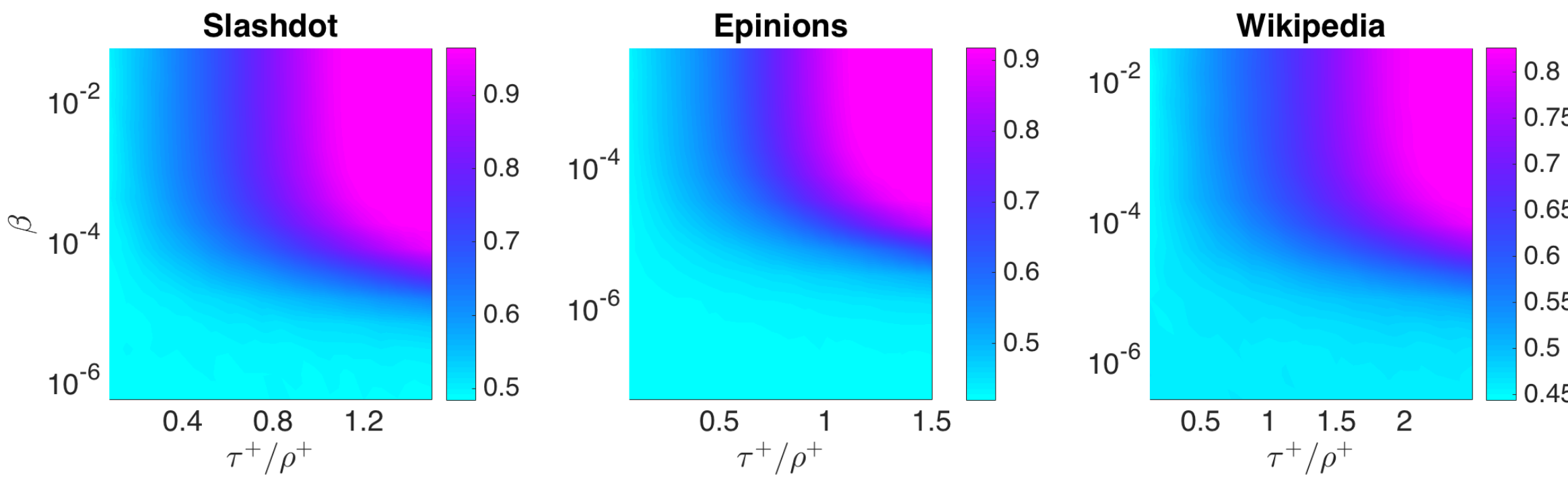}
\caption{{\bf Positive reciprocity bias}. The upper panels show the ratio between the average contribution to reputation from unreciprocated positive ratings $\lambda_\Phi^+$ measured under our null assumption of random link rewiring, and its value in the empirical networks. Lower panels show the ratio between the average contribution to reputation from reciprocated positive ratings $\lambda_\Gamma^+$ measured under the null assumption and its value in the empirical networks. In each plot such ratio is shown as a function of the intensity of choice parameter $\beta$ and the reciprocity target $\tau^+$ (normalised by the positive reciprocity $\rho^+$ of the empirical networks). As it can be seen, in the empirical networks the contribution to reputation from unreciprocated (reciprocated) positive links is systematically under-expressed (over-expressed) with respect to the null hypothesis.}
\label{fig:lambda_plus_app}
\end{figure*}

Fig. \ref{fig:lambda_plus_app} shows the contributions to reputation obtained by averaging over large samples of the first null model (NM1) as functions of both the intensity of choice parameter $\beta$ and the reciprocity target $\tau^+$. As it can be seen, we systematically find that the average contribution to reputation from unreciprocated positive links $\lambda_\Phi^+$ is under-expressed with respect to any null hypothesis, and, symmetrically, we find that the contribution from reciprocated positive links $\lambda_\Gamma^+$ is over-expressed. As in Fig. 2, $\lambda_\Phi^+$ has a non monotonic behaviour as a function of the reciprocity target, and it reaches a maximum whose value depends on $\beta$, whereas $\lambda_\Gamma^+$ always displays a monotonically non-decreasing behaviour both as a function of $\beta$ and $\tau^+$.
\begin{figure*}[h!]
\centering
\includegraphics[width=\textwidth]{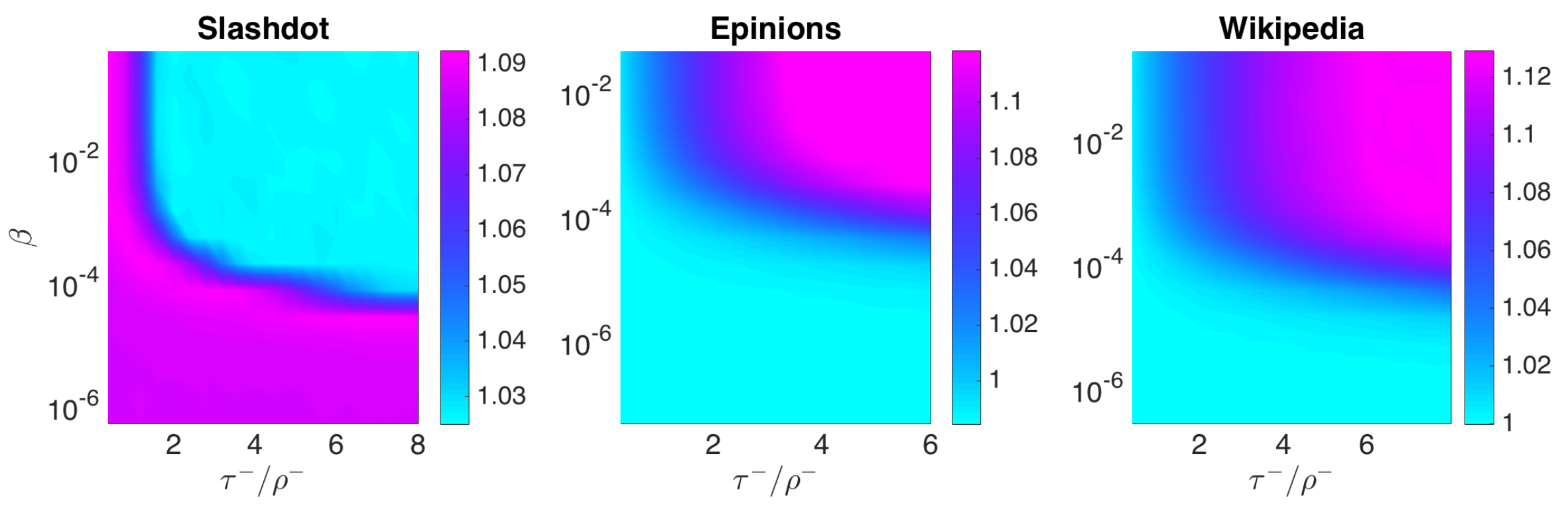}
\includegraphics[width=\textwidth]{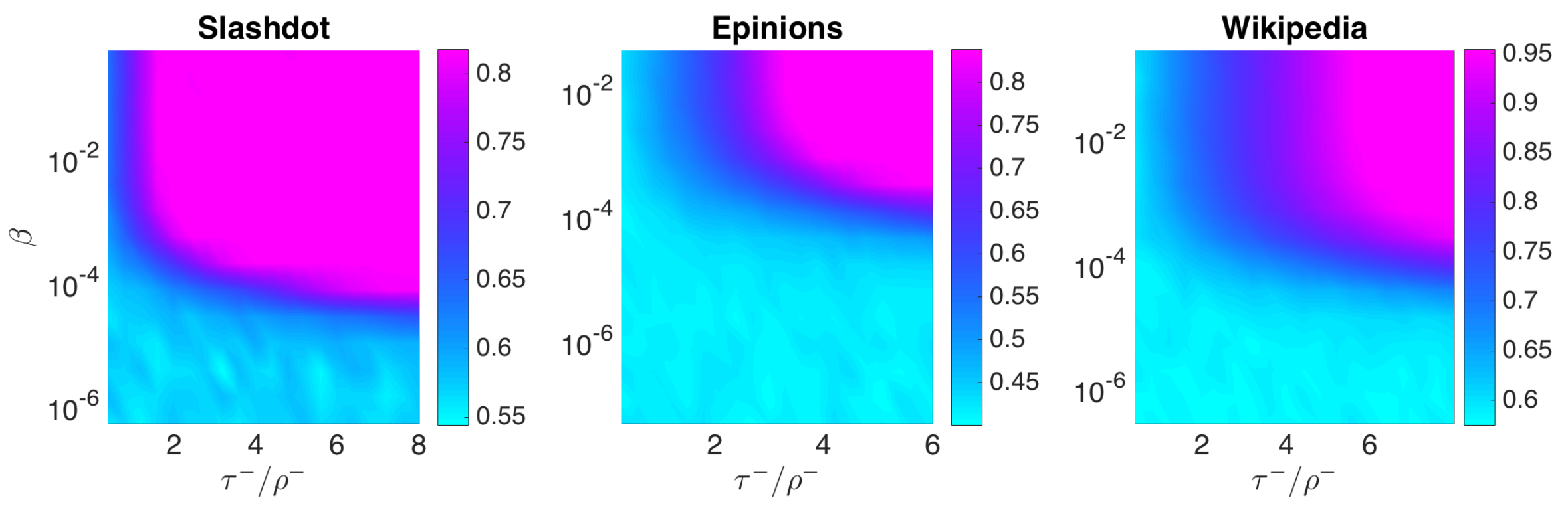}
\caption{{\bf Negative reciprocity bias}. The upper panels show the ratio between the average contribution to reputation from unreciprocated negative ratings $\lambda_\Phi^-$ measured under our null assumption of random link rewiring and its value in the empirical networks. Lower panels show the ratio between the average contribution to reputation from reciprocated negative ratings $\lambda_\Gamma^-$ measured under the null assumption and its value in the empirical networks. In each plot such ratio is shown as a function of the intensity of choice parameter $\beta$ and the reciprocity target $\tau^-$ (normalised by the negative reciprocity $\rho^-$ of the empirical networks). As it can be seen, in the empirical networks the contribution to reputation from unreciprocated (reciprocated) negative links is systematically under-expressed (over-expressed) with respect to the null hypothesis.}
\label{fig:lambda_minus}
\end{figure*}

Fig. \ref{fig:lambda_minus} shows the dependence of the average contributions to reputation from negative ratings. In analogy with the previous case, the contribution from unreciprocated links is systematically under-expressed while that from reciprocated links is systematically over-expressed. In this case, however, Slashdot displays a behaviour which markedly differs from the one observed in Epinions and Wikipedia. In fact, in the latter networks both $\lambda_\Phi^-$ and $\lambda_\Gamma^-$ are monotonically increasing functions of $\tau^-$, which shows that increasing the negative reciprocity target increases both retaliation and constructive negative feedback. Conversely, $\lambda_\Phi^-$ in Slashdot decreases as a function of the target $\tau^-$, i.e. the negative contribution to reputation from constructive feedback decreases as reciprocity is increased. This is suggestive of a different signature of genuinely hostile negative interactions, and suggests that reciprocity should be systematically discouraged in polarised environments. 

\section{Statistical properties of preference similarity in null model 2}
\label{nm2}

In the main paper we distinguish between two null models in order to assess the statistical significance of our results compared to a null hypothesis that does not preserve a proxy of homophily in the network, as opposed to a null hypothesis that does. We label this latter as null model 2 (NM2), and the quantity it preserves (together with the reputation of each individual user) is the preference similarity between two nodes, which for a pair of nodes $(i,j)$ reads $S_{ij} = \sum_{\ell=1}^N A_{i\ell} A_{j\ell}$. Such a null assumption enforces an additional constraint on the rewiring procedure we employ to build and sample our null models, which is specifically designed to incorporate the empirically observed level of agreement between pairs or users in the platform into our null models. 

The rationale behind the above assumption is that users who reciprocate can be reasonably expected to agree more in their approval/disapproval of other peers with respect to users who do not reciprocate. In order to test whether this is indeed the case we compute the preference similarity between pairs of nodes that share a positive reciprocated relationship (i.e. pairs $(i,j)$ such that $A_{ij} = A_{ji} = +1$), and compare its statistical properties to the preference similarity measured between pairs of nodes who share a positive unreciprocated relationship (i.e. pairs $(i,j)$ such that $A_{ij} = +1$ and $A_{ji} = 0$). We label the former quantity as $S^\leftrightarrow$ and the latter as $S^\rightarrow$. In Tables \ref{tab:moments1} and \ref{tab:moments2} we report the mean, variance, skewness and kurtosis of those two quantities as computed over all pertinent pairs of nodes in the three platforms we analyse.

As it can be seen, in all three networks $S^\leftrightarrow$ has a markedly larger mean than $S^\rightarrow$, which signals that indeed, on average, users who reciprocate tend to agree more than users who do not. However, both the distributions of $S^\leftrightarrow$ and $S^\rightarrow$ are significantly leptokurtic and skewed to the right, due to the presence of very active pairs of users with strong preference similarity. Interestingly, this behaviour is much more pronounced in the distributions of $S^\rightarrow$, meaning that the largest outliers in preference similarity are observed between pairs of nodes who do not reciprocate.

\begin{figure}[h!]
\centering
\includegraphics[width=1.0\columnwidth]{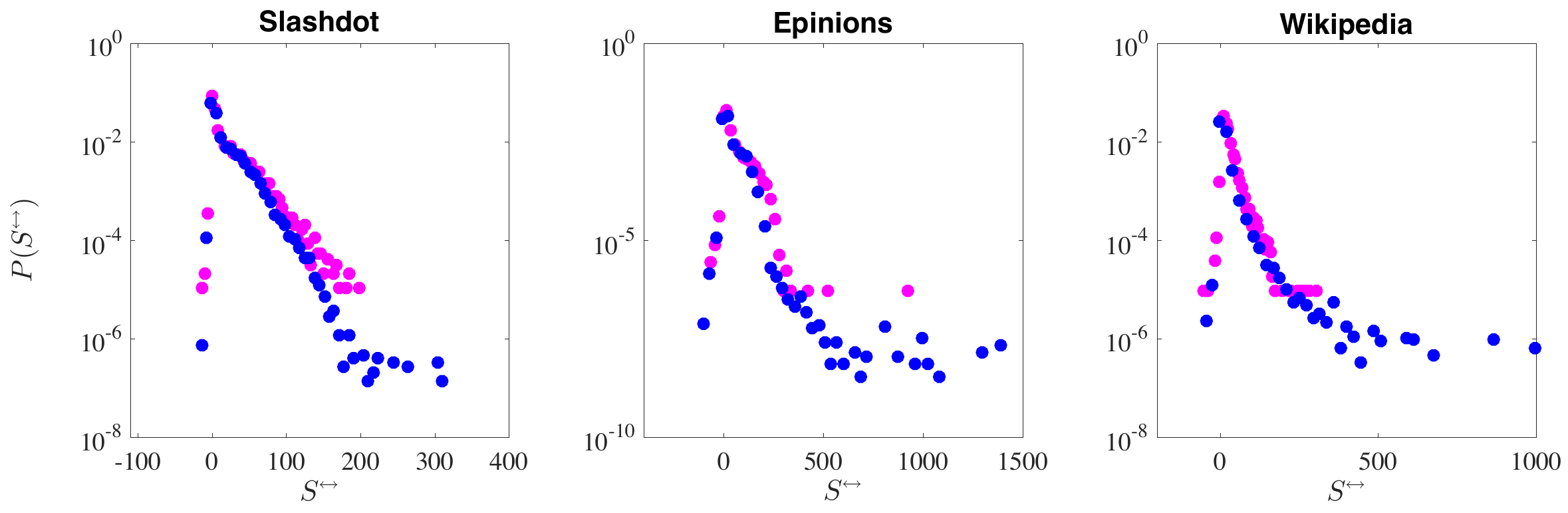}
\includegraphics[width=0.97\columnwidth]{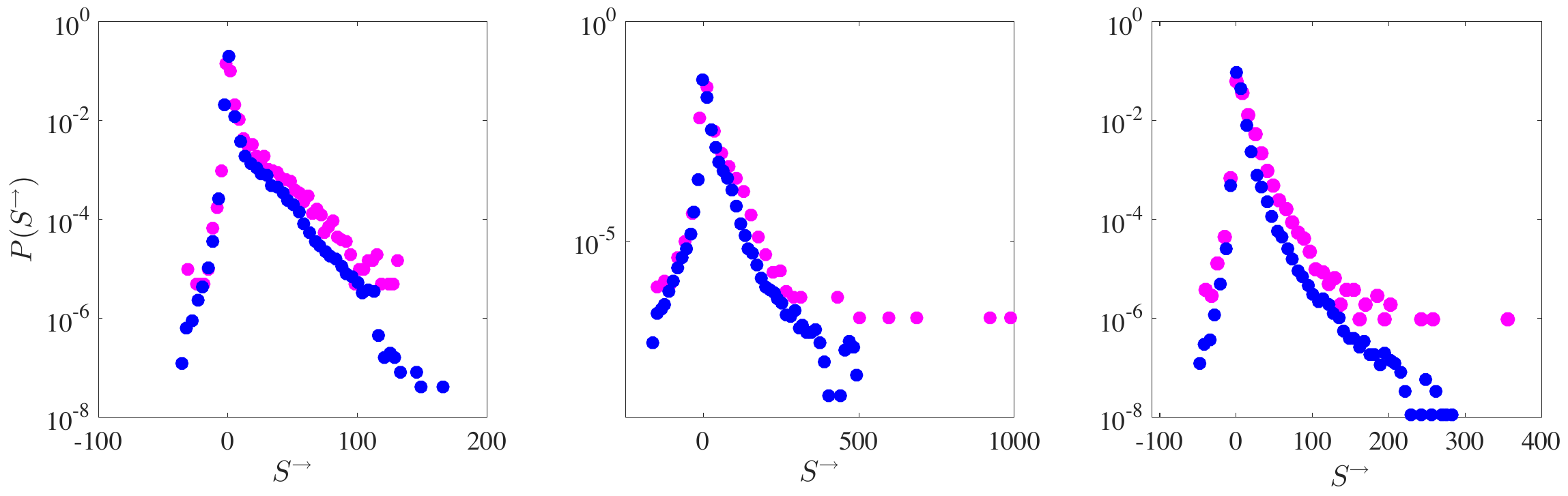}
\caption{\textbf{Distribution of preference similarity.} Distributions of the preference similarity $S^\leftrightarrow$ between pairs of nodes that share a positive reciprocated relationship (top panels) and of the preference similarity $S^\rightarrow$ between pairs of nodes that share one unreciprocated relationship. In all plots blue dots refer to the empirical networks, while pink dots refer to results obtained by averaging over NM2.}
\label{fig:ps_distr}
\end{figure}

As an additional robustness check, we test whether the distributions of $S^\leftrightarrow$ and $S^\rightarrow$ are compatible with those observed under NM2 (at positive reciprocity targets kept equal to the networks' empirical reciprocity, i.e. $\tau^+ = \rho^+$). In Fig. \ref{fig:ps_distr} we plot visual comparisons of the empirical distributions of such quantities and the ones obtained by sampling configurations of NM2, and in Tables \ref{tab:moments1} and \ref{tab:moments2} we compare the empirical moments with the corresponding $99\%$ significance level intervals obtained with NM2 (we mark with an asterisk the empirical moments that are compatible with such intervals).
\begin{table}[h!]
\centering
\begin{tabular}{|l|c|c|c|c|c|c|c|c|}
\hline
& \multicolumn{4}{c|}{Empirical networks} & \multicolumn{4}{c|}{Null model 2} \\
\hline
& Mean & Variance & Skewness & Kurtosis & Mean & Variance & Skewness & Kurtosis \\
\hline
Slashdot & $14.2$ & $4.84 \times 10^2$ & $2.26^*$ & $9.17^*$ & $[11.5; 11.7]$ & $[3.31; 3.44] \times 10^2$ & $[2.22; 2.41]$ & 
$[8.73; 11.86]$ \\
\hline
Epinions & $31.0$ & $1.95 \times 10^3$ & $2.36^*$ & $10.1^*$ & $[23.2; 23.5]$ & $[1.13; 1.16] \times 10^3$ & $[2.18; 2.88]$ & 
$[7.7; 37.7]$ \\
\hline
Wikipedia & $20.2$ & $3.89 \times 10^{2*}$ & $3.07$ & $22.0$ & $[10.8; 11.4]$ & $[2.86; 4.89] \times 10^2$ & $[6.6; 16.0]$ & $[0.89; 5.36] \times 10^2$ \\
\hline
\end{tabular}
\caption{\label{tab:moments1} {\bf Statistical properties of $S^\leftrightarrow$.} Moments of the distribution of $S^\leftrightarrow$ across all pairs of nodes that share a positive reciprocated relationship in the empirical networks, and $99\%$ confidence level intervals of the same moments computed in NM2 (for $\tau^+ = \rho^+$). Empirical moments marked with an asterisk are those that are compatible with the corresponding confidence level interval in the null models.}
\end{table}
\begin{table}[h!]
\centering
\begin{tabular}{|l|c|c|c|c|c|c|c|c|}
\hline
& \multicolumn{4}{c|}{Empirical networks} & \multicolumn{4}{c|}{Null model 2} \\
\hline
& Mean & Variance & Skewness & Kurtosis & Mean & Variance & Skewness & Kurtosis \\
\hline
Slashdot & $3.11$ & $72.2$ & $5.11^*$ & $38.2^*$ & $[0.42; 3.04]$ & $[3.0; 63.1]$ & $[4.72; 11.42]$ & $[0.33; 2.67] \times 10^2$ \\
\hline
Epinions & $11.30$ & $3.64 \times 10^2$ & $4.88$ & $87.4$ & $[4.68; 7.52]$ & $[0.74; 2.48] \times 10^2$ & $[3.19; 4.86]$ & $[24.7; 55.8]$ \\
\hline
Wikipedia & $7.15$ & $94.4$ & $3.93^*$ & $46.4^*$ & $[1.03; 5.80]$ & $[3.2; 60.6]$ & $[2.39; 6.21]$ & $[13.4; 79.7]$ \\
\hline
\end{tabular}
\caption{\label{tab:moments2} {\bf Statistical properties of $S^\rightarrow$.} Moments of the distribution of $S^\rightarrow$ across all pairs of nodes that have share an unreciprocated positive relationship in the empirical networks, and $99\%$ confidence level intervals of the same moments computed in NM2 (for $\tau^+ = \rho^+$). Empirical moments marked with an asterisk are those that are compatible with the corresponding confidence level interval in the null models.}
\end{table}

As it can be seen, NM2 generally underestimates the empirical means, but still preserves the empirically observed fact that the mean in the distribution of $S^\leftrightarrow$ is significantly larger than the one of $S^\rightarrow$ (indeed, in all three networks the $99\%$ confidence level intervals for the two means are mutually incompatible). Analogously, with the exception of $S^\leftrightarrow$ in Wikipedia, NM2 systematically underestimates the empirical variances. 

Higher order moments present a more heterogeneous picture. In fact, NM2 preserves the distributional properties of Slashdot in very good detail, as both the skewness and kurtosis of $S^\leftrightarrow$ and $S^\rightarrow$ are compatible with the empirically measured ones. On the other hand, NM2 captures well the distributional properties of $S^\leftrightarrow$ in Epinions while overestimating the skewness and kurtosis of the corresponding quantity in Wikipedia. Symmetrically, it captures well the properties of $S^\rightarrow$ in Wikipedia while underestimating the skewness and kurtosis in Epinions.

All in all, it can be concluded that NM2 succeeds at preserving some non-trivial trends and properties observed in the empirical networks, while also preserving at least the correct order of magnitudes of the quantities it does not fully capture in a statistical sense.

\section{A null model based on link and sign reshuffling}
\label{link_signs}

As a further check of the robustness of our results, we test the statistical significance of the  features observed in the empirical networks against a null hypothesis based on random link and sign reshuffling. Namely, we perform the same rewiring operations carried out for null model 1 (top-left picture in Figure 1) on the unsigned networks where each entry is taken equal to $|A_{ij}|$, and later randomly reassign $+/-$ signs on the links in the same proportions as they occur in the empirical networks. This procedure randomises the network topology while preserving its overall heterogeneity in terms of the unsigned degree of each node $k_i = \phi^+ + \phi^- + \gamma^+ + \gamma^-$ (see Eqs. (1) and (2)), which corresponds to the overall number of ratings received by each user. By reassigning signs at the end of the random rewiring procedure we therefore build network configurations where all correlations between positive/negative links are removed (except for those due to the higher density of positive links). 
\begin{table}[h!]
\centering
\begin{tabular}{|l|c|c|c|c|c|c|}
\hline
& \multicolumn{6}{c|}{Empirical networks} \\
\hline
& $\rho^+$ & $\rho^-$ & $\lambda_\Phi^+$ & $\lambda_\Gamma^+$ & $\lambda_\Phi^-$ & $\lambda_\Gamma^-$ \\
\hline
Slashdot & $41.3\%$ & $15.9\%$ & $0.0323$ & $0.0355$ & $0.0340$ & $0.0522$ \\
\hline
Epinions & $42.4\%$ & $7.70\%$ & $0.0156$ & $0.0220$ & $0.0236$ & $0.0157$  \\
\hline
Wikipedia & $17.6\%$ & $8.50\%$ & $0.0270$ & $0.0325$ & $0.0362$ & $0.0321$ \\
\hline
& \multicolumn{6}{c|}{Null model} \\
\hline
Slashdot & $[1.58; 1.82]\%$ & $[0.32; 0.66]\%$ & $[0.0343; 0.0347]$ & $[0.0176; 0.0207]$ & $[0.0336; 0.0350]$ & $[0.0148; 0.0286]$ \\
\hline
Epinions & $[2.58; 2.75]\%$ & $[0.17; 0.34]\%$ & $[0.0189; 0.0190]$ & $[0.0085; 0.0091]$ & $[0.0184; 0.0191]$ & $[0.0069; 0.0105]$ \\
\hline
Wikipedia & $[2.77; 3.05]\%$ & $[0.36; 0.66]\%$ & $[0.0295; 0.0297]$ & $[0.0139; 0.0155]$ & $[0.0286; 0.0297]$ & $[0.0099; 0.0197]$ \\
\hline
\end{tabular}
\caption{\label{tab:null_signs} {\bf Results from a null model based on link and sign reshuffling.} In the first four rows we report, for the sake of convenience, the values of positive and negative reciprocity $\rho^\pm$ and the average contributions to reputation from reciprocated ($\lambda_\Gamma^\pm$) and unreciprocated ($\lambda_\Phi^\pm$) links measured in the empirical networks, while in the next four rows we report the $99\%$ confidence level intervals for the corresponding quantities measured under a null assumption of random link and sign reshuffling.}
\end{table}

Table \ref{tab:null_signs} summarises our findings in such null model. Namely, we report the values of positive/negative reciprocity $\rho^\pm$ and the contributions to reputation from reciprocated ($\lambda_\Gamma^\pm$) and unreciprocated ($\lambda_\Phi^\pm$) links as measured in the empirical networks (whose values are also reported in Tables 1, 2, and 3) and in the aforementioned null model. From a qualitative point of view, in most cases we observe the same results we obtained when analysing the results from the two null models discussed in the main paper. Namely:
\begin{itemize}
\item In the empirical networks both positive and negative reciprocity are markedly over-expressed (i.e. by more than one order of magnitude) with respect to the null assumption.
\item The contribution to reputation from positive unreciprocated (reciprocated) activity $\lambda_\Phi^+$ ($\lambda_\Gamma^+$) in the empirical networks is under-expressed (over-expressed) with respect to the null assumption.
\item Whereas in all three networks we analyse we have $\lambda_\Gamma^+ > \lambda_\Phi^+$, i.e. on average one reciprocated link contributes to reputation more than one unreciprocated link, under the above null assumption we observe the opposite relationship.
\item The average contribution to reputation from reciprocated negative links ($\lambda_\Gamma^-$) in the empirical networks is systematically over-expressed with respect to the above null assumption.
\end{itemize} 

Altogether, the above points confirm that the reciprocity bias persists against a null assumption of random link and sign reshuffling in the positive case. On the other hand, it should be noted that the contribution to reputation from unreciprocated negative links ($\lambda_\Phi^-$) measured in the empirical networks is compatible with the one measured under the above null assumption in the case of Slashdot, and over-expressed in the case of Epinions and Wikipedia. This is at odds with the observations made in NM1 (see Appendix \ref{gen_nm1}), and indicates that a random redistribution of the ratings either preserves or underestimates the contribution to reputation from unilateral negative feedback.

\section{Link elimination}
\label{link_el}

One of our main results concerns the fragility of the states observed in real-life P2P networks. Indeed, as we show in the main paper, the random elimination of a small fraction (i.e. $3 - 11\%$ depending on the network) of reciprocated positive links is sufficient to remove the reciprocity bias and make the contribution to reputation from positive unreciprocated ratings ($\lambda_\Phi^+$) and reciprocated ratings ($\lambda_\Gamma^+$) statistically compatible. Let us remark that the protocol chosen to carry out the rating elimination procedure is crucially important to keep the fraction of removed ratings so low. Indeed, the above efficiency is achieved only when selecting \emph{pairs of users} at random. Eliminating reciprocated ratings at random requires instead the elimination of a much larger number of ratings, as it can be seen in Fig. \ref{fig:link_removal}. Given the heavy tailed nature of the distributions of ratings given and received by each node (see Fig. (\ref{fig:ratings_distr})), this means that the most efficient way to decrease the contribution to reputation from reciprocated activity is to eliminate the reciprocated ratings between users with a lower number of ratings. In contrast, removing any rating with equal probability amounts to preferentially removing ratings between high activity users, i.e. hubs in the P2P network. We discuss in the main paper how this result is meaningful from the viewpoint of user incentives.
\begin{figure*}[h!]
\centering
\includegraphics[width=\textwidth]{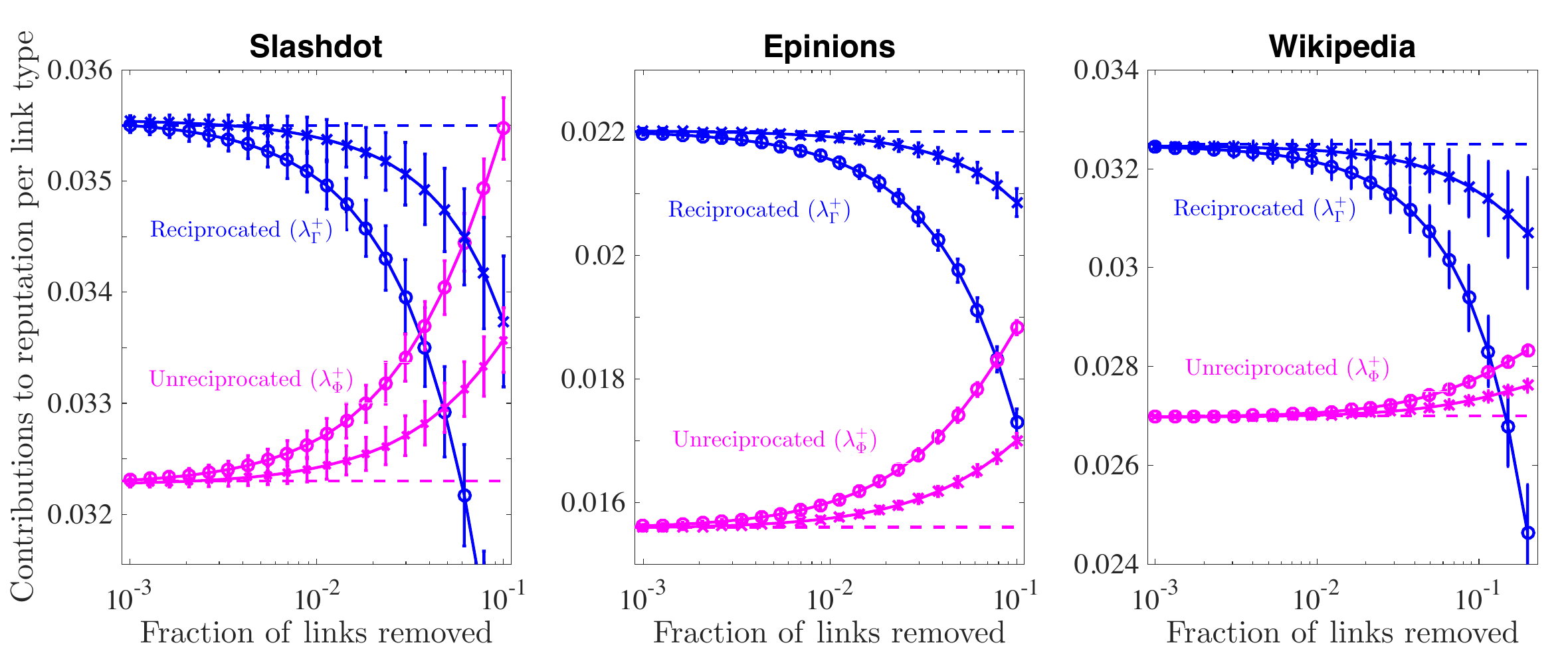}
\caption{{\bf Comparison between the performance of different link elimination protocols in removing the reciprocity bias.} Solid lines show the average contribution to reputation from unreciprocated ($\lambda_\Phi^+$, pink) and reciprocated ($\lambda_\Gamma^+$, blue) positive links as a function of the fraction of reciprocated positive links removed from the network. Circles represent the behaviour of such quantities when a random node selection protocol is followed, i.e. nodes are chosen at random with uniform probability and reciprocated positive links between them, if any, are removed (this case corresponds to the one shown in Fig. 3). Crosses refer instead to a random link selection protocol, where links are removed with uniform probability. In the former case the majority of links removed are between low degree nodes, whereas in the latter case the elimination procedure targets hubs with higher probability. The dashed lines represent the values of $\lambda^+_\Gamma$ (upper line) and $\lambda_\Phi^+$ (lower line) in the original networks. Error bars represent $99\%$ confidence level intervals.}
\label{fig:link_removal_app}
\end{figure*}

\section{Robustness with respect to different thresholds}
\label{robust}

All the results in our paper have been obtained by restricting empirical networks to a high participation core of actively engaged users with at least $t = 10$ ratings (either given or received). In this Appendix we provide evidence that all the main results we discuss in the main paper are robust with respect to changes in the threshold $t$. In particular, we discuss the case of an \emph{extended participation core} to show the effects of a lower threshold ($t=5$), which allows to take into account the contribution of less actively engaged users. In Table VIII we report the number of nodes and positive/negative links for such networks. As it can be seen, consistently with the case $t=10$, we still observe a large prevalence of positive links, which represent $80\%$ or more of the total links. The only slight qualitative difference with respect to the networks used in the main paper is an increase in sparsity, which is due to the inclusion of a substantial number of nodes with low activity.
\begin{table}[h!]
\begin{center}
\begin{tabular}{|l|c|c|c|c|c|}
\hline
& $N$ & $L^+$ & $L^-$ & $\xi^+$ & $\xi^-$ \\
\hline
Slashdot & $8,982$ & $159,861$ & $47,437$ & $2.0 \times 10^{-3}$ & $5.9 \times 10^{-4}$ \cr
\hline
Epinions & $14,192$ & $496,545$ & $52,643$ & $2.5 \times 10^{-3}$ & $2.6 \times 10^{-4}$ \cr
\hline
Wikipedia & $8,995$ & $214,692$ & $37,121$ & $2.7 \times 10^{-3}$ & $4.6 \times 10^{-4}$ \cr
\hline
\end{tabular}
\caption{{\bf Network statistics in the extended participation core.} Number of users $N$, number of positive ($L^+$) and negative ($L^-$) ratings, and sparsity $\xi^\pm$ in the participation core obtained for $t = 5$.}
\end{center}
\label{tab:ext_core_network_stats}
\end{table} 

\begin{table}[h!]
\centering
\begin{tabular}{|l|c|c|c|c|c|}
\hline
& & \multicolumn{2}{c|}{Null model 1} & \multicolumn{2}{c|}{Null model 2} \\
\hline
& $\rho^+$ & $\rho^+_0$ & $\rho^+_\mathrm{SAT}$ & $\rho^+_0$ & $\rho^+_\mathrm{SAT}$ \\
\hline
Slashdot & $35.4\%$ & $[1.23; 1.46]\%$ & $[39.8; 40.0]\%$ & $[3.08; 3.18]\%$ & $[36.2; 36.4]\%$ \cr
\hline
Epinions & $41.0\%$ & $[1.89; 1.99]\%$ & $[46.3; 46.4]\%$ & $[3.31; 3.35]\%$ & $[41.5; 41.6]\%$ \cr
\hline
Wikipedia & $15.3\%$ & $[2.39; 2.55]\%$ & $[32.2; 32.5]\%$ & $[2.33; 2.39]\%$ & $[21.7; 22.0]\%$ \cr
\hline
\end{tabular}
\caption{\label{tab:pos_reciprocity_t5} {\bf Over-expression of positive reciprocity in the extended participation cores.} Comparison between the positive reciprocity $\rho^+$ observed in the extended participation cores of the three networks we analyse and the $99\%$ confidence level intervals for the corresponding ``basal'' levels $\rho_0^+$ and saturation levels $\rho^+_\mathrm{SAT}$ obtained under a null hypothesis of random link rewiring constrained to preserve each user's reputation (null model 1), and a null hypothesis further constrained to also preserve the preference similarity of each pair of nodes (null model 2).}
\end{table}

\begin{table}[h!]
\centering
\begin{tabular}{|l|c|c|c|c|c|}
\hline
& & \multicolumn{2}{c|}{Null model 1} & \multicolumn{2}{c|}{Null model 2} \\
\hline
& $\rho^-$ & $\rho^-_0$ & $\rho^-_\mathrm{SAT}$ & $\rho^-_0$ & $\rho^-_\mathrm{SAT}$ \\
\hline
Slashdot & $13.8\%$ & $[0.21; 0.40]\%$ & $[21.2; 21.8]\%$ & $[6.52; 6.89]\%$ & $[16.8; 17.2]\%$ \cr
\hline
Epinions & $6.88\%$ & $[0.80; 1.10]\%$ & $[22.0; 22.5]\%$ & $[5.13; 5.55]\%$ & $[16.0; 16.5]\%$ \cr
\hline
Wikipedia & $7.38\%$ & $[1.55; 1.95]\%$ & $[45.0; 45.6]\%$ & $[6.93; 7.90]\%$ & $[30.7; 31.2]\%$ \cr
\hline
\end{tabular}
\caption{\label{tab:neg_reciprocity_t5} {\bf Over-expression of negative reciprocity in the extended participation cores.} Comparison between the negative reciprocity $\rho^-$ observed in the extended participation cores of the three networks we analyse and the $99\%$ confidence level intervals for the corresponding ``basal'' levels $\rho_0^-$ and saturation levels $\rho^-_\mathrm{SAT}$ obtained under a null hypothesis of random link rewiring constrained to preserve each user's reputation (null model 1), and a null hypothesis further constrained to also preserve the preference similarity of each pair of nodes (null model 2).}
\label{tab:rep_production}
\end{table}

In Tables \ref{tab:pos_reciprocity_t5} and \ref{tab:neg_reciprocity_t5} we summarise the values of positive/negative reciprocity $\rho^\pm$ we measure in the participation cores for $t = 5$, as well as the corresponding basal reciprocity $\rho_0^\pm$ and saturation reciprocity $\rho_\mathrm{SAT}^\pm$ we obtain by averaging over the two classes of null models we consider with $\beta = 0$ and $\beta \rightarrow \infty$, respectively. By comparing such values in the $t = 10$ and $t = 5$ cases one can notice that essentially the reciprocity levels observed in the empirical networks and their relative magnitude with respect to those computed over null model samples remain qualitatively very similar. As a consequence of this fact, we still observe a substantial over-expression of reciprocity, especially in the positive case, with respect to the basal levels $\rho_0^\pm$, and we still observe that Slashdot and Epinions display positive reciprocity close to their respective saturation levels $\rho_\mathrm{SAT}^\pm$ (especially those computed under NM2), whereas Wikipedia's positive and negative reciprocity remain quite far from the corresponding saturation levels.

In full analogy with the results reported in the main paper, we find that in the extended participation cores the average contribution $\lambda_\Gamma^+$ to reputation from positive reciprocated ratings is systematically higher than the average contribution $\lambda_\Phi^+$ from unreciprocated positive ones. Also in analogy with the restricted participation core analysed in the main paper, we again observe that only in Slashdot the contribution to reputation from reciprocated negative ratings exceeds that from unreciprocated negative ones (i.e. $\lambda_\Gamma^- > \lambda_\Phi^-$). 
\begin{table}[h!]
\centering
\begin{tabular}{|l|c|c|c|c|}
\hline
& $\lambda_\Phi^+$ & $\lambda_\Gamma^+$ & $\lambda_\Phi^-$ & $\lambda_\Gamma^-$  \cr
\hline
Slashdot & $0.0391$ & $0.0504$ & $0.0411$ & $0.0632$ \cr
\hline
Epinions & $0.0207$ & $0.0311$ & $0.0344$ & $0.0251$ \cr
\hline
Wikipedia & $0.0328$ & $0.0397$ & $0.0469$ & $0.0409$ \cr
\hline
\end{tabular}
\label{tab:lambda_ext_core}
\caption{{\bf Evidence that reciprocated ratings contribute more to reputation than unreciprocated ones in the extended participation core.} Average contribution to reputation from each link category: $\lambda_\Phi^\pm$ denote the average contribution from a positive/negative unreciprocated rating, while $\lambda_\Gamma^\pm$ denote the average contribution from a positive/negative reciprocated rating.}
\end{table} 

The above consistency of results is also detected in terms of reciprocity bias. In the main paper (i.e. on the $t=10$ participation core) we observe the following two features:
\begin{itemize}
\item A systematic over-expression of the average contribution to reputation from positive/negative reciprocated ratings ($\lambda_\Gamma^\pm$) in the empirical networks with respect to any null hypothesis (i.e. for any value of $\beta$ or $\tau^\pm$) preserving the networks' reputation landscape and the networks' local preference similarity structure.
\item For large values of $\beta$ (i.e. when the rewiring procedure is made very selective), the average contribution to reputation from positive unreciprocated ratings in the null models is larger than the average contribution from reciprocated positive ratings (i.e., $\lambda_\Phi^+ > \lambda_\Gamma^+$) over a wide range of reciprocity targets $\tau^+$, as opposed to what is observed in the empirical networks.
\end{itemize}

\begin{figure*}[h!]
\centering
\includegraphics[width=\textwidth]{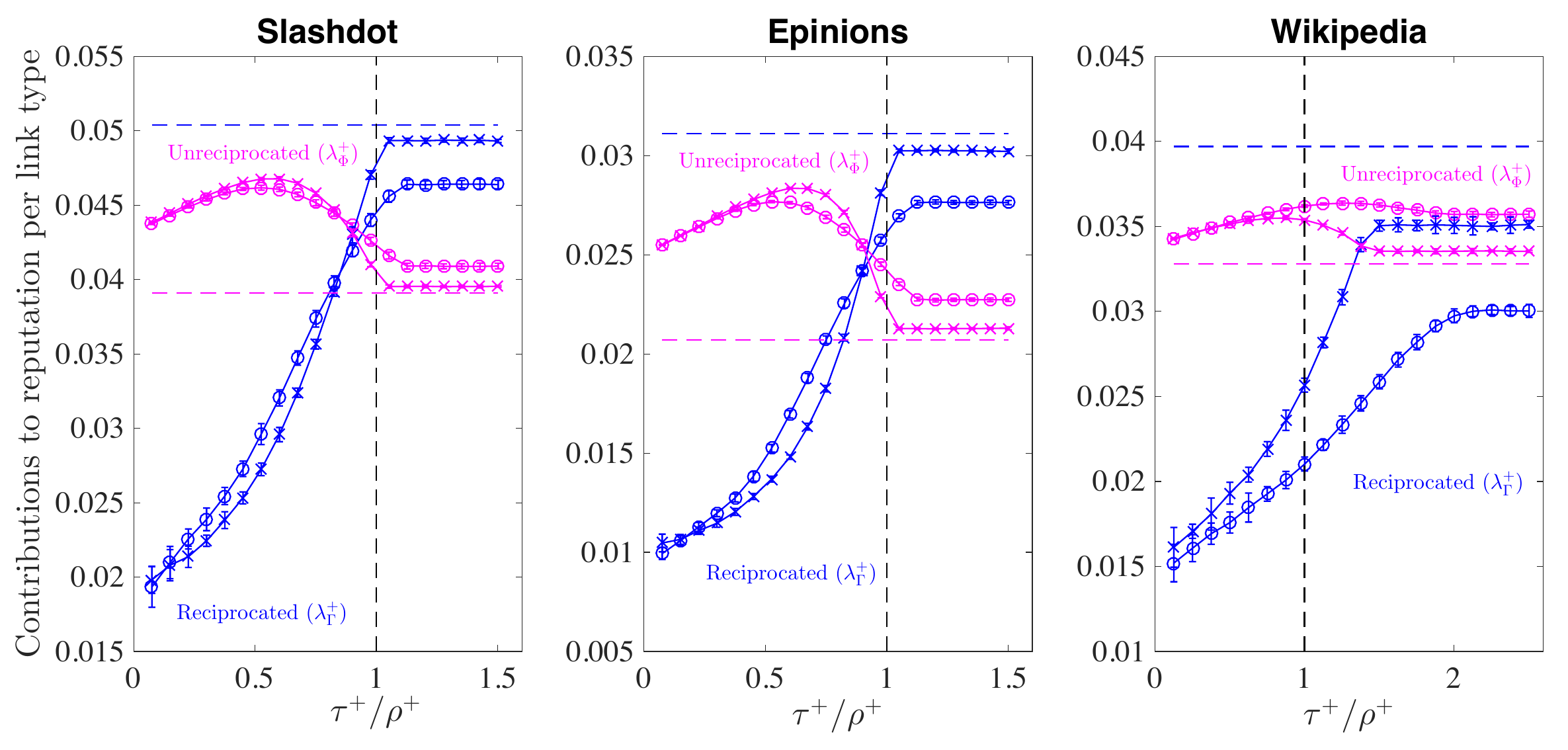}
\caption{{\bf Reciprocity bias in the extended participation core}. Behaviour of the average contribution to reputation from unreciprocated positive ratings ($\lambda_\Phi^+$, pink) and reciprocated positive ratings ($\lambda_\Gamma^+$, blue) under two null hypotheses of random link rewiring designed to produce a predefined positive reciprocity target $\rho^+$ in the extended participation core. Circles refer to a null hypothesis constrained to preserve the reputation of each user (null model 1), while crosses refer to a null hypothesis further constrained to also preserve the preference similarity of each pair of nodes (null model 2). The behaviour of $\lambda_\Phi^+$ and $\lambda_\Gamma^+$ is shown as a function of the ratio between the reciprocity target $\tau^+$ and the positive reciprocity $\rho^+$ measured in the actual platforms (first column in Table \ref{tab:pos_reciprocity_t5}). Error bars correspond to $99\%$ confidence level intervals. Dashed lines correspond to the values of $\lambda_\Phi^+$ (pink) and $\lambda_\Gamma^+$ (blue) measured in the actual platforms (i.e. to the values reported in columns 1 and 2, respectively, of Table \ref{tab:lambda_ext_core}). The fact that the contribution from reciprocated (unreciprocated) activity in the actual platforms is systematically lower (higher) than under our null hypotheses highlights the existence of the reciprocity bias in the extended participation core.}
\label{fig:lambda_high_beta_t5}
\end{figure*}

Both the above points are illustrated for the $t=5$ participation core in Fig. \ref{fig:lambda_high_beta_t5}, which closely resembles Fig. 2.

\begin{figure*}[h!]
\centering
\includegraphics[width=\textwidth]{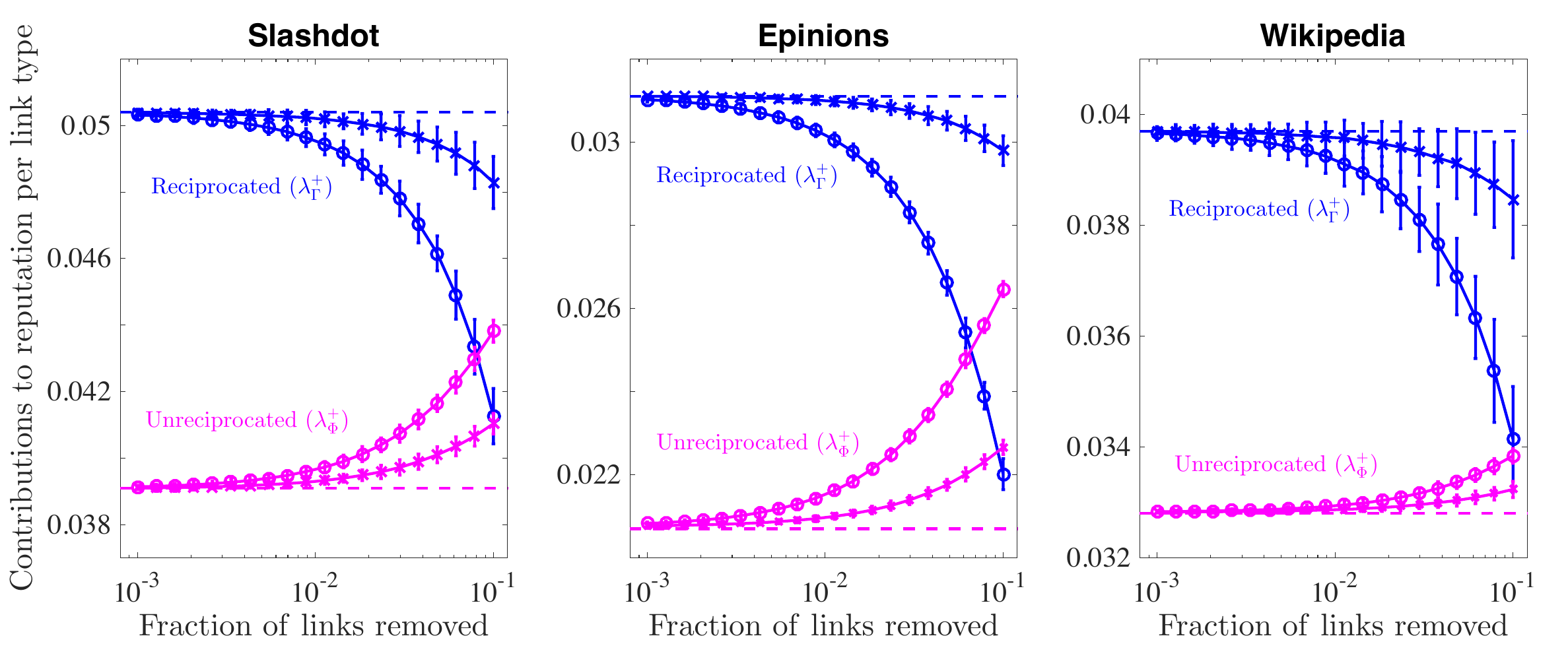}
\caption{{\bf Elimination of the reciprocity bias in the extended participation core}. Solid lines show the average contribution to reputation from unreciprocated ($\lambda_\Phi^+$, pink) and reciprocated ($\lambda_\Gamma^+$, blue) positive links as a function of the fraction of reciprocated positive links removed from the extended participation core network. Circles represent the behaviour of such quantities when a random node selection protocol is followed, i.e. nodes are chosen at random with uniform probability and reciprocated positive links between them, if any, are removed. Crosses refer instead to a random link selection protocol, where links are removed with uniform probability. In the former case the majority of links removed are between low degree nodes, whereas in the latter case the elimination procedure targets hubs with higher probability. The dashed lines represent the values of $\lambda^+_\Gamma$ (upper line) and $\lambda_\Phi^+$ (lower line) in the original networks. Error bars represent $99\%$ confidence level intervals.}
\label{fig:link_removal_t5}
\end{figure*}

As discussed in the main paper, we find the reciprocity bias to be a peculiar property of real-life P2P platforms. Indeed, small perturbations are enough to make the contributions to reputation from unreciprocated ($\lambda_\Phi^+$) and reciprocated ($\lambda_\Gamma^+$) positive ratings statistically compatible. We detect the same phenomenon in the extended participation core: As shown in Fig. \ref{fig:link_removal_t5}, the removal of a small fraction of reciprocated positive ratings is enough to first make $\lambda_\Phi^+$ and $\lambda_\Gamma^+$ statistically compatible, and to eventually make the former the prevalent contribution to reputation, as opposed to what is observed in the empirical networks. In particular, we find that the removal of $8\%$ randomly selected reciprocated positive ratings in Slashdot (which correspond to less than $3\%$ of the overall positive ratings) is enough to make $\lambda_\Phi^+$ and $\lambda_\Gamma^+$ statistically compatible. The same result is achieved by removing $6\%$ of the reciprocated positive ratings in Epinions (corresponding to $2.5\%$ of the overall positive ratings) and $10\%$ of the reciprocated positive ratings in Wikipedia (i.e. $1.5\%$ of the overall positive ratings).

In full analogy with the case discussed in Appendix \ref{link_el}, we find that the link elimination protocol adopted to remove the ratings plays a crucial role in keeping the overall fractions of removed links so low. Indeed, we find that the most efficient protocol is the one based on the random selection of pairs of nodes and the subsequent elimination of possible reciprocated ratings between them. On the other hand, as shown in Fig. \ref{fig:link_removal_t5} a protocol based on the random selection of links is much less efficient and entails the removal of substantial portions of links in order to achieve statistical compatibility between $\lambda^+_\Gamma$ and $\lambda_\Phi^+$.

\end{document}